\documentclass[twocolumn,showpacs,preprintnumbers,amsmath,amssymb]{revtex4}
\pdfoutput=1

\usepackage{graphicx}
\usepackage{dcolumn}
\usepackage{bm}
\usepackage {url}

\newcommand{\pb}{$^{208}\mathrm{Pb}^{82+}$}

\begin{document}


\title{Measurements of heavy ion beam losses from collimation}

\author{R.~Bruce}
 \altaffiliation[Also at ]{MAX-lab, Lund University, Sweden.}
 \email{roderik.bruce@cern.ch}
\author{R.W.~Assmann}
\author{G.~Bellodi}
\author{C.~Bracco}
\author{H.H.~Braun}
\author{S.~Gilardoni}
\author{E.B.~Holzer}
\author{J.M.~Jowett}
\author{S.~Redaelli}
\author{T.~Weiler}
\affiliation{CERN, Geneva, Switzerland}

\date{\today}

\begin{abstract}

The collimation efficiency for $^{208}$Pb$^{82+}$ ion beams in the LHC is predicted to be lower than requirements. Nuclear fragmentation and electromagnetic dissociation in the primary collimators create fragments with a wide range of $Z/A$ ratios, which are not intercepted by the secondary collimators but lost where the dispersion has grown sufficiently large. In this article we present measurements and simulations of loss patterns generated by a prototype LHC collimator in the CERN SPS. Measurements were performed at two different energies and angles of the collimator. We also compare with proton loss maps and find a qualitative difference between $^{208}$Pb$^{82+}$ ions and protons, with the maximum loss rate observed at different places in the ring. This behavior was predicted by simulations and provides a valuable benchmark of our understanding of ion beam losses caused by collimation.

\end{abstract}

\maketitle

\section{INTRODUCTION}
The Large Hadron Collider (LHC)~\cite{lhcdesignV1}, presently being commissioned at CERN, will collide beams of protons and later heavy nuclei~\cite{jowett08}, starting with \pb, at energies never reached before. The main parameters of the beams of the two species are listed in Tab.~\ref{tab:LHCbeams}. The LHC uses superconducting magnets, which operate with a high magnetic field at 8.33~T, near the quench limit, meaning that even a small temperature rise, of the order of a 1~K, can make the magnets pass from a superconducting to a resistive state. At the same time, with a stored proton beam energy of 362~MJ (see Tab.~\ref{tab:LHCbeams}), even a small beam loss of $4\times10^7$ protons in a magnetic element might induce enough heating to cause a quench. Therefore, all beam losses need to be tightly controlled and, for this purpose, a collimation system has been designed~\cite{lhcdesignV1,assmann04,assmann05,assmann06}. This system is located in two warm insertions of the LHC and cleans halo particles from the beam in a controlled manner before they are lost elsewhere.

The LHC collimation system is primarily optimized for proton operation but will be used also during ion runs. The collimation is however predicted to be less efficient for ions than for protons~\cite{braun04}, although the stored beam energy is almost a factor~100 lower, as can be seen in Tab.~\ref{tab:LHCbeams}. In order to understand why this is the case, we give a very brief summary of the physics processes occurring when particles traverse the collimator material in Sec.~\ref{sec:coll-physics} and then of the functioning of the collimation system in Sec.~\ref{sec:lhc-ion-coll}.

To better understand the LHC ion collimation, we have made an experiment on ion collimation in the CERN Super Proton Synchrotron (SPS), which is the main subject of this article. The experimental setup is presented in Sec.~\ref{sec:exp-setup} and the simulation methods in Sec.~\ref{sec:icosim}. In the remaining sections, the measurement results are compared to simulations and we also make a brief comparison with the case of protons.

\section{PHYSICS OF HEAVY IONS IN COLLIMATORS}
\label{sec:coll-physics}
In order to understand the LHC ion collimation inefficiency, we give a brief overview of the physical processes occurring when heavy nuclei traverse the collimator material.

A short review of the passage of charged particles through matter can be found in Ref.~\cite{pdg} and an extensive theoretical treatment of ions in particular in Ref.~\cite{eichler90}. Here we highlight two important processes: the energy loss through ionization, which is described by the well-known Bethe-Bloch formula, and the change of direction through many small-angle scattering events, so-called multiple Coulomb scattering (MCS). Angular deviations can also be caused by nuclear elastic scattering, which is a significant effect for protons but negligible for \pb ions.

\begin{figure}[tb]
 \includegraphics[width=6.5cm]{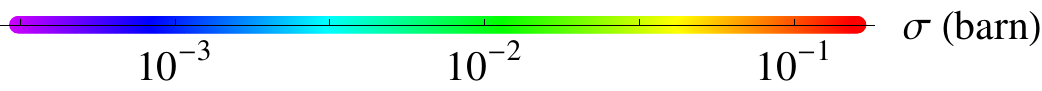}\\
 \includegraphics[width=8.5cm]{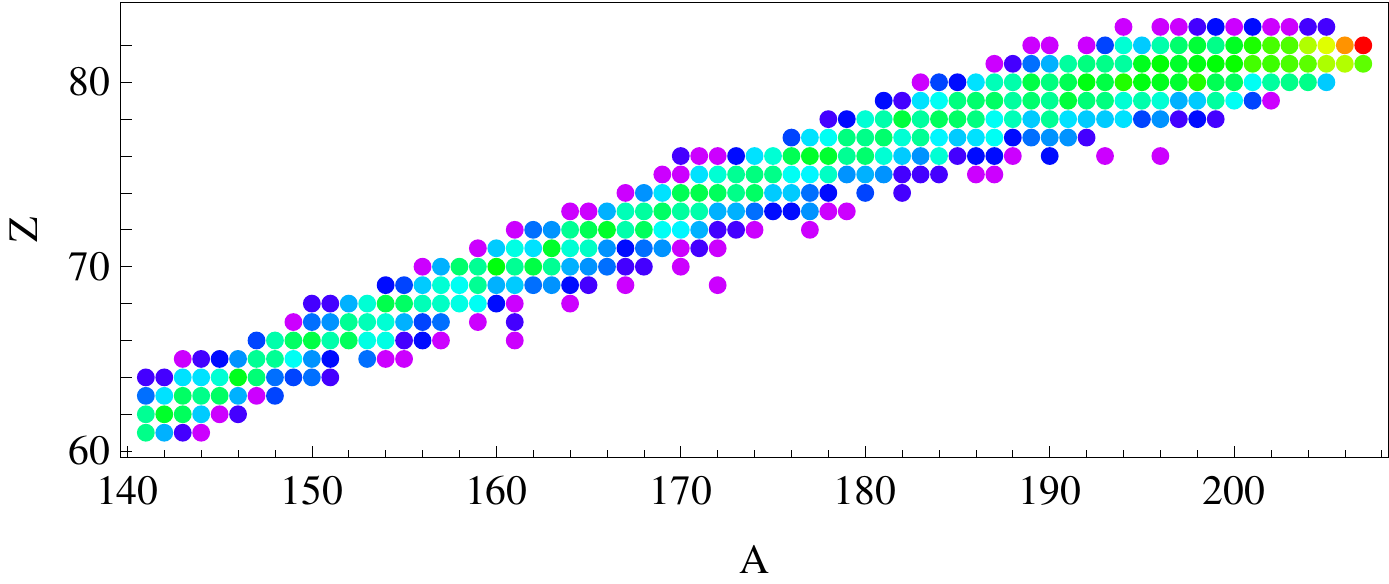}
 \caption{The cross sections for creating heavy isotopes (mass number $A>140$ and charge $Z>60$) by a 106.4~GeV/nucleon \pb beam impacting on a carbon target. The cross sections were simulated by FLUKA.}
 \label{fig:frag-XS}
\end{figure}

These processes are present for all charged particles, while some others are peculiar to heavy ions. An impinging nucleus may lose one or several nucleons, in particular neutrons, through electromagnetic dissociation (EMD), which is a process with a logarithmic energy dependence taking place in ultraperipheral collisions. For a review of this field, see Refs.~\cite{baur98,bert02,bert05}. The nuclei may also split up in smaller fragments through nuclear inelastic reactions (see Ref.~\cite{scheidenberger04} for measurements of nuclear fragmentation of \pb ions at 158~GeV/nucleon in a carbon target, a situation close to the SPS collimation experiment described in this article). The cross section for nuclear inelastic interactions is only weakly dependent on energy and the created fragments have a wide range of masses. In Fig.~\ref{fig:frag-XS} we show the total cross sections for the creation of different isotopes, the sum of the electromagnetic and hadronic part, by 106.4~GeV/nucleon \pb ions in a carbon target, as calculated by the Monte Carlo program FLUKA~\cite{fluka1,fluka2,roes00}. This corresponds to the situation in the collimator in the SPS experiment.

In each interaction, a number of secondary particles are created that constitute the hadronic shower~\cite{ferarri96}. The final energy deposition in the material is to a large extent due to these secondaries.

\section{LHC ion collimation}
\label{sec:lhc-ion-coll}
The LHC collimation system consists of primary and secondary collimators~\cite{jeanneret98}, both made of graphite jaws. Particles at large amplitudes hit first a primary collimator, which is shorter and the limiting aperture of the machine. There they interact in such a way that they are either absorbed or scattered back into the beam or to an even larger amplitude. The latter particles form a secondary halo, which is intercepted and absorbed by the secondary collimators, possibly several turns later. The condition on the angular kick $\Delta x'$ given by the primary collimator for a particle to be intercepted by the secondary collimator is~\cite{jeanneret98}
\begin{equation}
\Delta x'>\sqrt{\frac{(N_2^2-N_1^2)\epsilon_N}{\gamma \beta_x}}.
\label{eq:collKick}
\end{equation}
Here $\epsilon_N$ is the normalized emittance, $\gamma$ the usual Lorentz factor, $\beta_x$ the usual optical function at the primary collimator and $N_1,N_2$ are the transverse positions of the primary and secondary collimators, given in the number of $\sigma$ (the standard deviation of an assumed Gaussian beam distribution) from the center of the vacuum chamber.

\begin{table} \centering
  \caption{The parameters for the p$^+$ and \pb LHC beams at collision.}
  \label{tab:LHCbeams}
  \begin{ruledtabular}
\begin{tabular}{|l|r|r|}
  Particle & p$^+$ & \pb \\ \hline \hline
  Energy/nucleon & 7 TeV & 2.759 TeV \\ \hline
  Relativistic $\gamma$ & 7461 & 2963.5 \\ \hline
  No. bunches & 2808 & 592 \\ \hline
  No. particles/bunch & 1.15$\times10^{11}$ & $7\times10^7$ \\ \hline
  Transverse normalized & & \\
  RMS emittances & 3.75 $\mu$m & 1.5 $\mu$m \\ \hline
  Stored beam energy & 362 MJ & 3.81 MJ \\ \hline
  Peak luminosity & $10^{34}$ cm$^{-2}$ s$^{-1}$ & $10^{27}$ cm$^{-2}$ s$^{-1}$ \\
\end{tabular}
\end{ruledtabular}
\end{table}

We can now understand the inefficiency of the LHC ion collimation by comparing the length of material that an ion would need to traverse in order to hit the secondary collimator with the interaction length of nuclear fragmentation processes. Taking standard LHC \pb parameters in Eq.~(\ref{eq:collKick}) ($N_1=6$, $N_2=7$, $\epsilon_N=1.5\;\mu$m, $\gamma_\mathrm{rel}=2963.5$, $ \beta_x\approx 135$~m), we see that the angular kick that a \pb ion would need to receive in the primary collimator in order to hit the secondary collimators is approximately 7~$\mathrm{\mu}$rad. To have this RMS angle of the ion distribution exiting a collimator, the required length of the jaw would be 2~m if we use the Gaussian approximation of the Moliere formula~\cite{pdg} to calculate the RMS MCS angle $\theta_0$:
\begin{equation}
\theta_0=\frac{13.6\;\mathrm{MeV}}{v p} Z_0 \sqrt{\frac{s}{S_0}} \left[1+0.038 \ln\left(\frac{s}{S_0}\right)\right]
\end{equation}
Here $Z_0$ is the charge number of the incident ion, $p$ its momentum, $v$ its speed, $s$ the distance traversed in the material and $S_0$ the radiation length.

However, the length of the primary collimators is only 60~cm, and typical trajectories inside them much shorter, while the interaction lengths for nuclear inelastic interactions and EMD are around 2.5~cm and 19~cm respectively~\cite{braun04}. Since the transverse recoil of the heavy fragments is very small, it is clear that the heavy ions that are not absorbed in the primary collimator are likely to be fragmented without having obtained a sufficiently large angle to reach the secondary collimators.

The created fragments (e.g. $^{207}$Pb, $^{203}$Tl and others) have a different magnetic rigidity $B\rho\,(1+\delta)$ from the main beam (magnetic rigidity is defined as $p/Ze=B\rho$ for an ion with momentum $p$ and charge $Ze$ that would have a bending radius $\rho$ in a magnetic field $B$). Therefore they are bent and focused differently. The fractional rigidity deviation $\delta$ is given by
\begin{equation}
\label{eq:delta-eff}
\delta=\frac{Z_0}{A_0}\frac{A}{Z}\left(1+\frac{\Delta p}{p_0}\right)-1,
\end{equation}
where $(Z_0,A_0)$ are the charge and mass number of the original ion, $(Z,A)$ of the fragment and $\Delta p/p_0$ the fractional momentum deviation per nucleon of the fragment with respect to the main beam.
These ions follow the locally generated dispersion function $d_x$ from the collimator and may be lost downstream in the machine where the horizontal aperture $A_x$ satisfies
\begin{equation}
d_x \, \delta \ge A_x.
\end{equation}
Here we have assumed that the dispersive contribution to the orbit is much larger than the betatronic part. Because of the different $\delta$ of the created ion species, they are lost at different localized spots, making the machine act as a spectrometer. Some impact locations could be outside the warm collimation insertion, meaning a potential risk of quenching superconducting magnets.

To study the LHC ion collimation inefficiency, a series of simulation studies have been done~\cite{braun04,epac06}. Since a large fraction of the systematic error in those simulations comes from the generation and tracking of the fragmented ions, an experiment on ion collimation in the CERN Super Proton Synchrotron (SPS) has been performed and the results, presented in the following, are compared to simulations, not only in terms of loss locations but with the goal of reproducing the absolute value of the losses measured by the beam loss monitors (BLMs). We also make a brief comparison with the case of protons.

\section{EXPERIMENTAL SETUP}
\label{sec:exp-setup}


The SPS is a synchrotron of 6.9 km circumference that now serves as the last injector for the LHC~\cite{lhcdesignV3} but also provides beam for various fixed-target experiments. It accelerates protons up to 450~GeV and \pb nuclei up to 177~GeV/nucleon before extraction. 

\begin{figure}[tb]
  \includegraphics*[width=8.5cm]{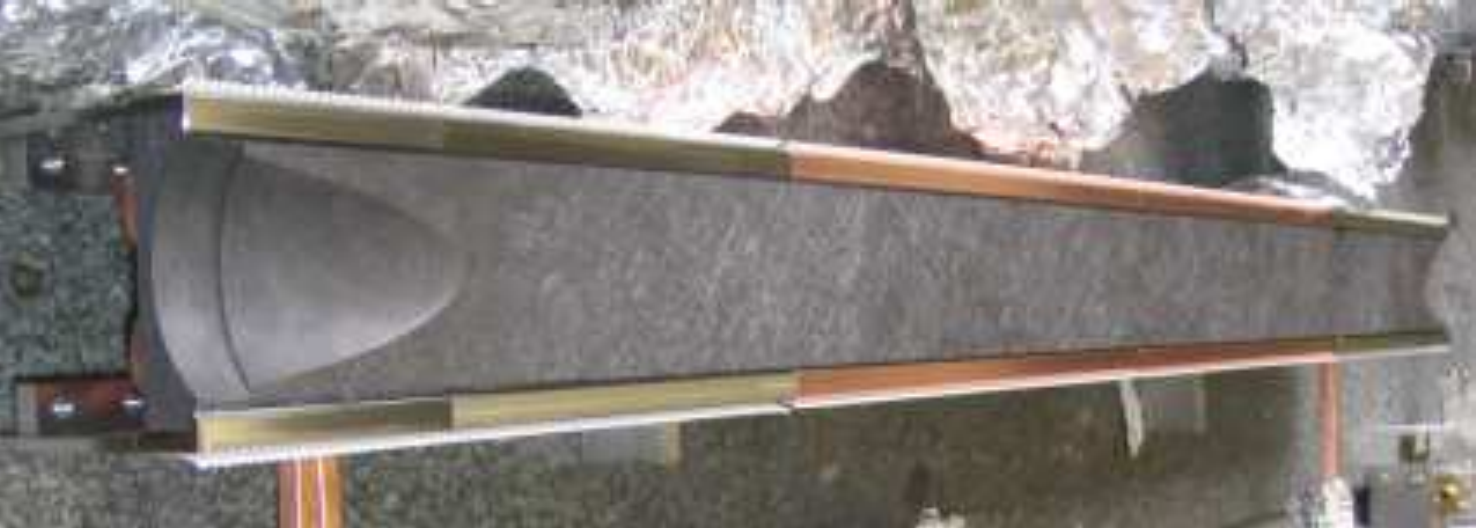}\\
  \caption{A collimator jaw of the same type as installed in the SPS. The figure is taken from~\cite{lhc-coll-web}.}
  \label{fig:sps-jaw}
\end{figure}

\begin{figure}[tb]
  \includegraphics*[width=8.5cm]{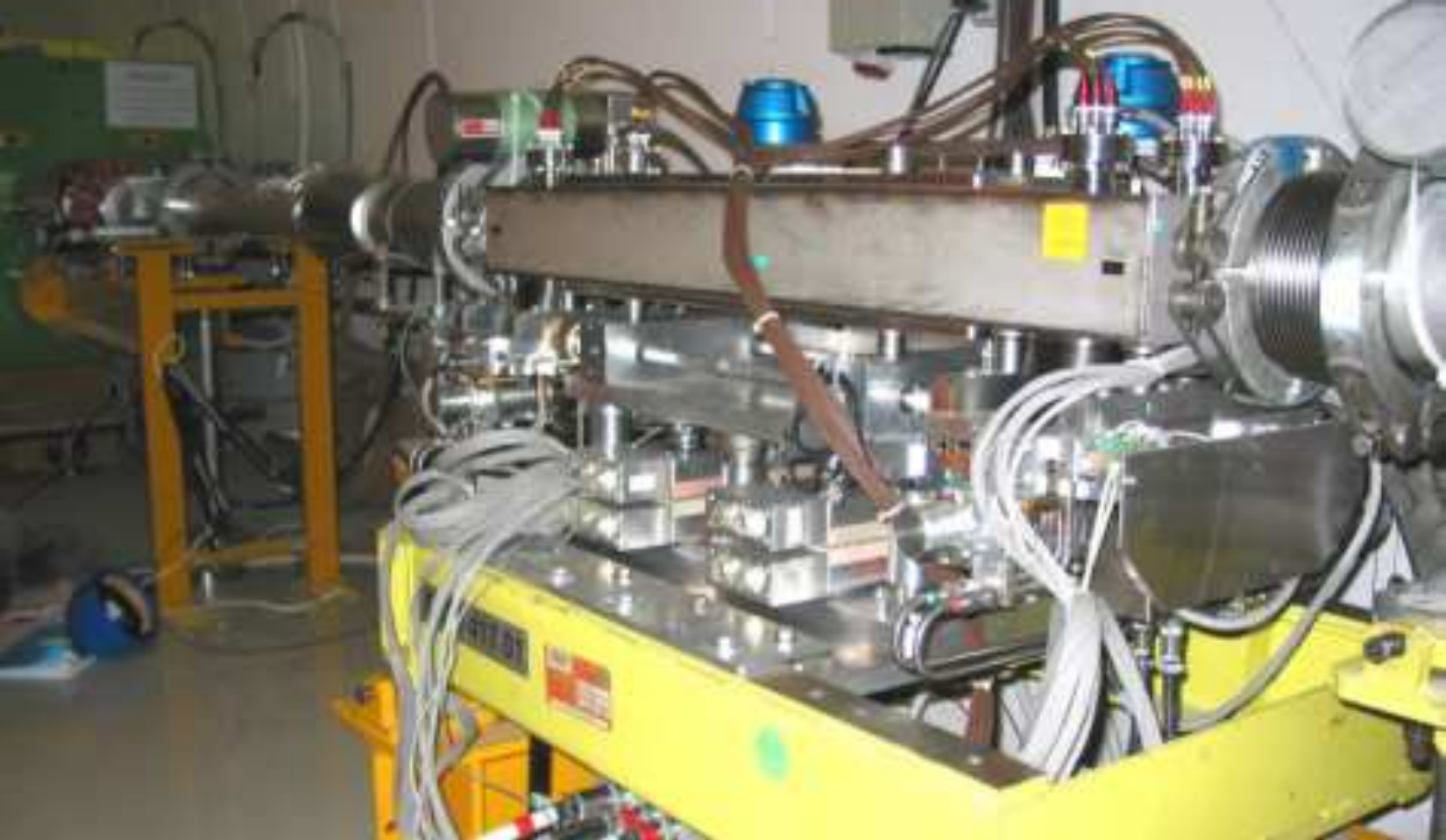}\\
  \caption{The installation of the prototype LHC collimator in the SPS tunnel. The figure is taken from~\cite{lhc-coll-web}.}
  \label{fig:sps-coll-installation}
\end{figure}

\begin{figure}[tb]
  \includegraphics*[width=9cm]{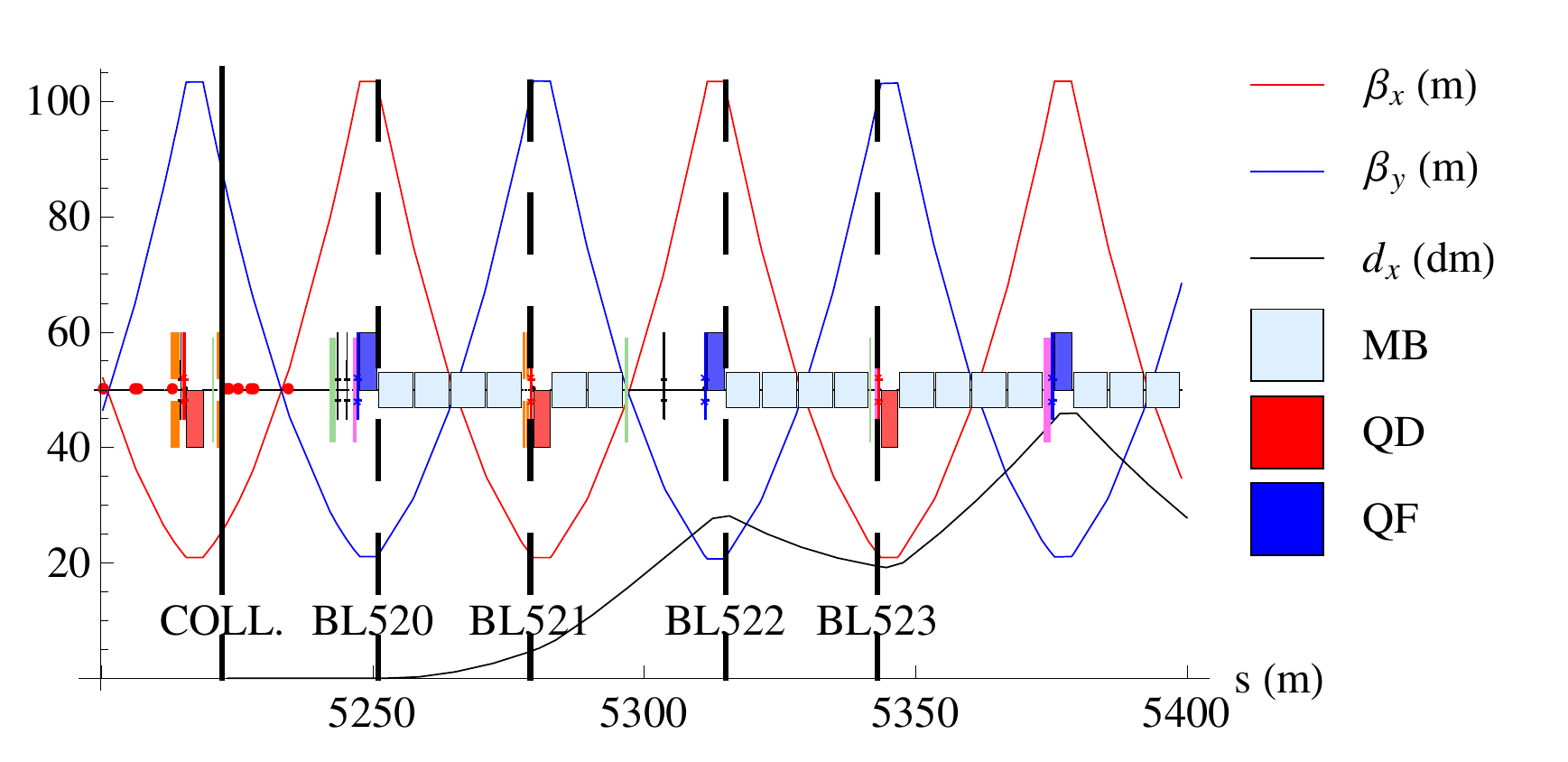}\\
  \caption{The $\beta$-functions of the SPS downstream of the collimator and the locally generated dispersion $d_x$ from the collimator. We show also the magnetic elements (MB=main bend, QF=focussing quadrupole, QD=defocussing quadrupole). The BLMs and the collimator are indicated by vertical lines.}
  \label{fig:sps-beamline}
\end{figure}

\begin{figure}[tb]
\centering
\includegraphics*[width=85mm]{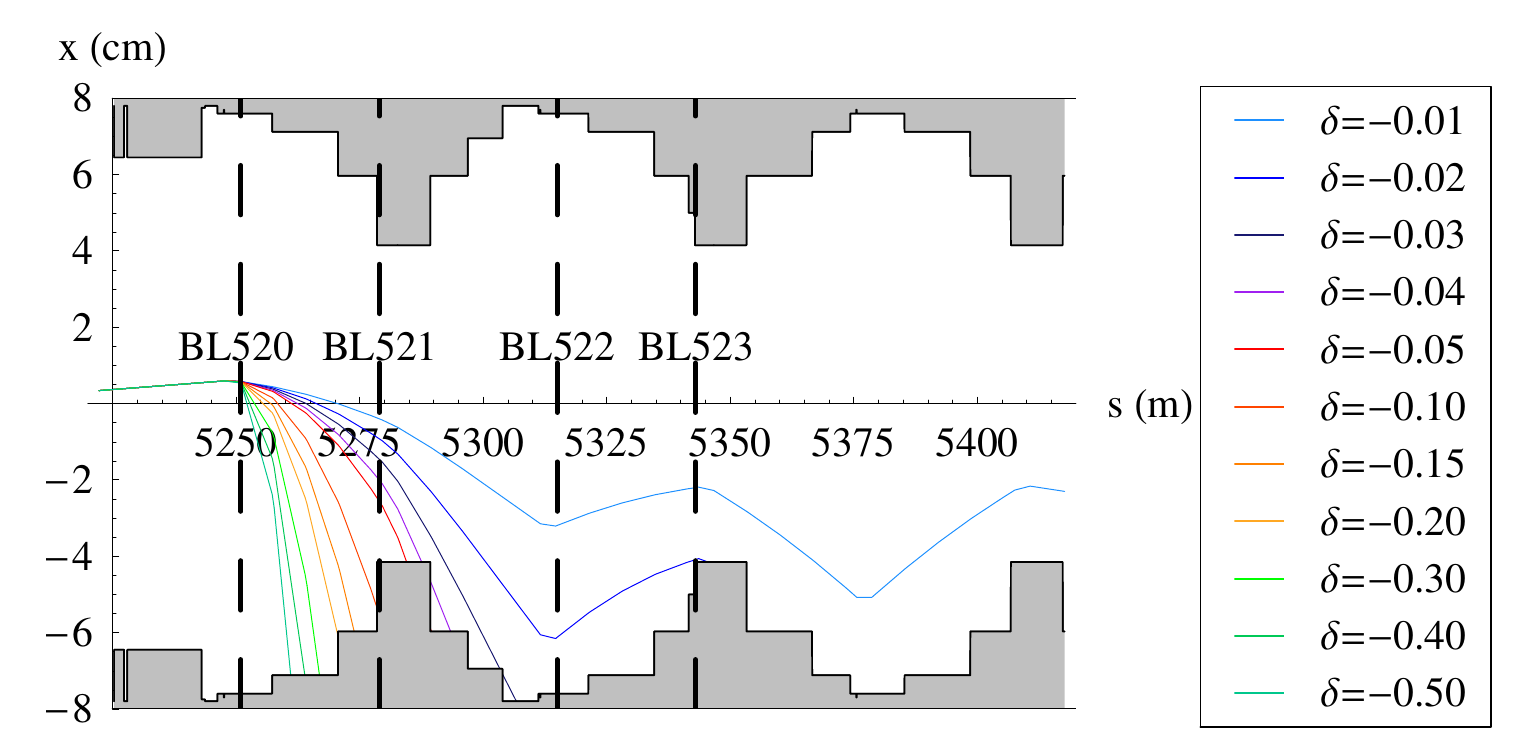}
\caption{Dispersive orbits of fragmented ions produced in one of the collimator jaws, shown together with the aperture. The collimator is located at $s=5222$~m at the left edge. The vertical dashed lines indicate the locations of the four BLMs closest downstream. A large fraction of the total losses occur at the aperture limitation at $s=5277$~m.}
\label{fig:ions-disp}
\end{figure}

\begin{figure}
  \includegraphics[width=85mm]{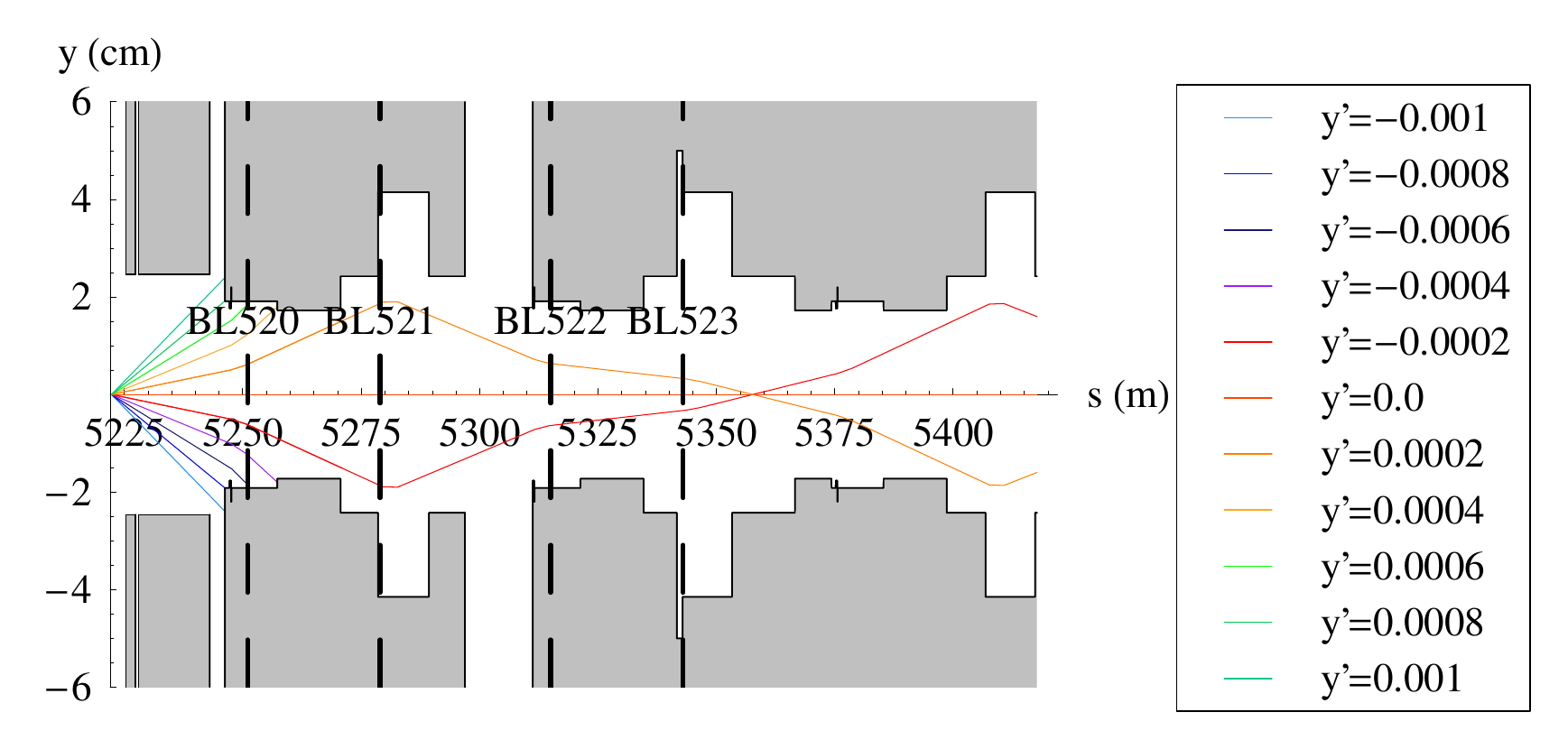}\\
  \caption{Vertical betatron orbits coming out of the collimator, starting at $y=0$, for different starting values of $y'$ (vertical momentum normalized by longitudinal momentum). The collimator is located at $s=5222$~m at the left edge. The vertical dashed lines indicate the location of the four BLMs closest downstream. Particles with large vertical angles are lost close to BL520.}
  \label{fig:vert-beta-orb}
\end{figure}

A prototype of a secondary LHC collimator has been installed in the long straight section~5 (LSS5)~\cite{bertarelli05, redaelli06}.  It consists of a pair of 1~m long CFC graphite jaws, which can be moved independently to intercept the beam in the horizontal plane. A similar jaw is shown in Fig.~\ref{fig:sps-jaw} and the installation in the SPS tunnel is shown in Fig.~\ref{fig:sps-coll-installation}. The collimator is located in a position with small horizontal $\beta$-function and dispersion in order not to introduce aperture bottlenecks. The optical functions and magnetic elements in the vicinity of the installation are shown in Fig.~\ref{fig:sps-beamline}, the horizontal aperture in Fig.~\ref{fig:ions-disp}, and the vertical aperture in Fig.~\ref{fig:vert-beta-orb}.

\begin{figure}[tb]
\centering
\includegraphics*[width=85mm]{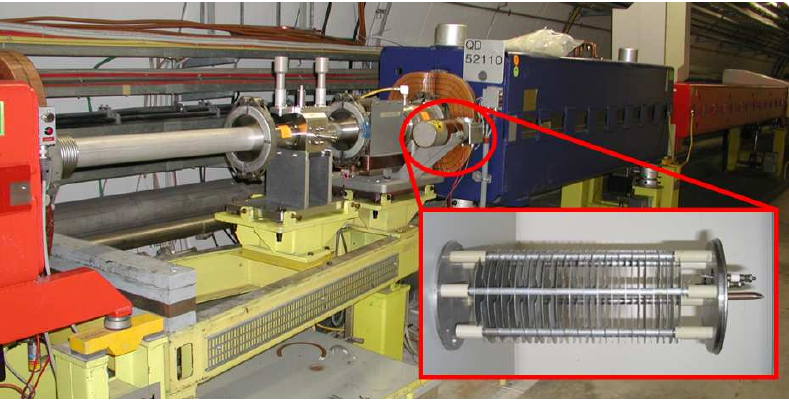}
\caption{The installation of a BLM (BL521) in the SPS tunnel, shown together with a closeup of the inner assembly of the chamber.}
\label{fig:sps-blm}
\end{figure}

Moving the collimator into the beam creates losses in the ring, since no second collimator is present to absorb scattered particles. There is however no risk for quenches, since the SPS does not use superconducting magnets. The losses are recorded by 216~BLMs placed around the machine~\cite{moy73}. The BLMs are ionization chambers mounted close to lattice quadrupoles (the inner part of a BLM and an example of the installation are shown in Fig.~\ref{fig:sps-blm}) and filled with N$_2$ gas at 1.1~bar. They consist of parallel aluminum plates, acting as anodes and cathodes. The losses are read out and integrated over every machine cycle (18~s for ions). Fig.~\ref{fig:sps-beamline} shows the $s$-values of the four BLMs (called BL520, BL521, BL522 and BL523) immediately downstream of the collimator. BL520 and BL522 are mounted in the vertical plane below the beamline, while BL521 and BL523 are in the horizontal plane on the inside of the ring.

 \begin{table*} \centering
  \caption{The parameters of the beams in the SPS used for the collimation experiment.}
  \label{tab:SPS-param}
  \begin{ruledtabular}
\begin{tabular}{|l|r|r|r|}
  Particle & \pb & \pb & p$^+$ \\ \hline
  Energy/nucleon (GeV) & 106.4 & 5.9 & 270  \\ \hline
  Injected intensity (particles)& 6--7$\times10^{7}$ & 5--9$\times10^7$ & $ 10^{12}$--$10^{13}$ \\ \hline
  Horizontal normalized emittance (1~$\sigma$)& 1 $\mu$m & 1 $\mu$m & 2.7 $\mu$m \\ \hline
  Vertical normalized emittance (1~$\sigma$)& 1 $\mu$m & 1 $\mu$m & 4 $\mu$m \\ \hline
  Collimator steps & 0.1-1 mm & fixed & 0.1-1 mm \\ \hline
  Collimator angles & 0 mrad, 2 mrad & 0 mrad & 0 mrad
\end{tabular}
\end{ruledtabular}
\end{table*}

\begin{figure}[tb]
\centering
\includegraphics*[width=75mm]{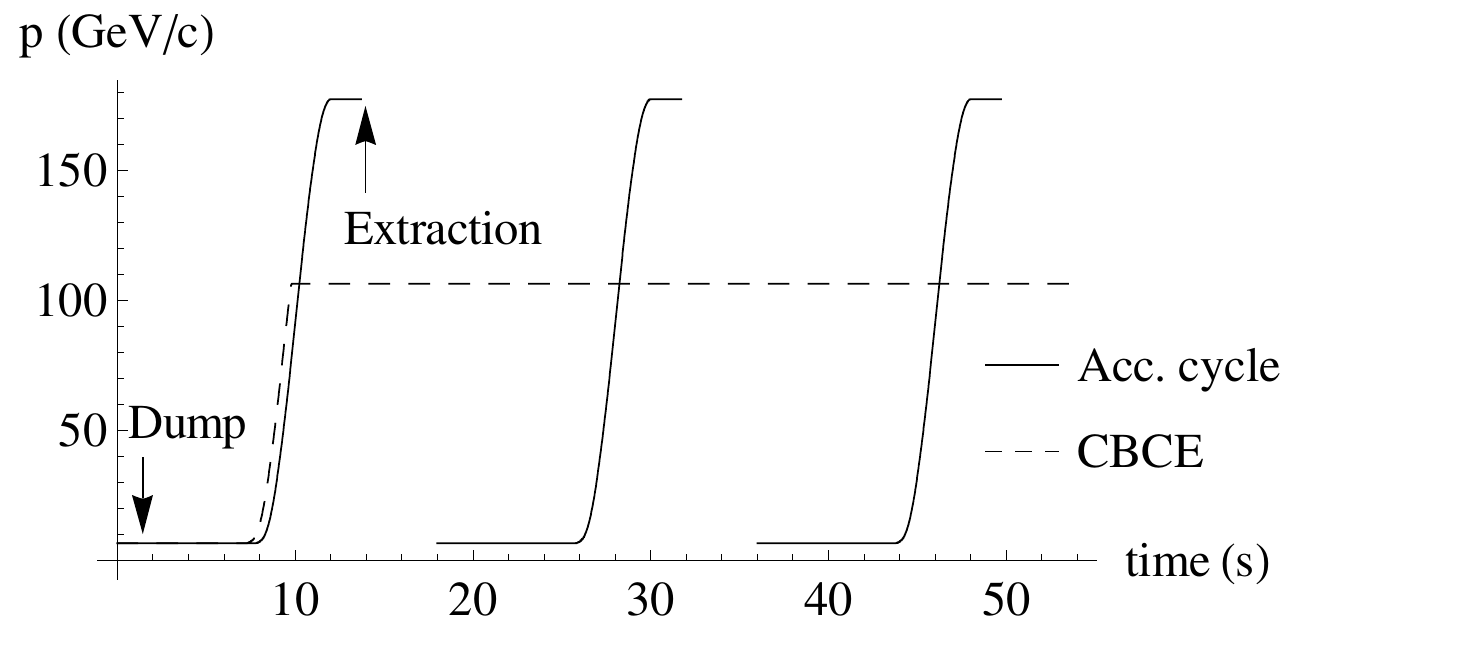}
\caption{The \pb momentum per nucleon during three cycles of the SPS in the case of acceleration and a circulating beam at constant energy (CBCE). In the acceleration cycle, the dumping of the beam during the collimation measurements is indicated, as well as the extraction momentum.}
\label{fig:mag-cycle}
\end{figure}

Beam loss data were collected during \pb dedicated ion runs in late 2007 and the main beam parameters are shown in Tab.~\ref{tab:SPS-param}. Data were taken both with a circulating beam at 106.4~GeV/nucleon and with a 5.9~GeV/nucleon beam intercepted by the collimator directly after injection. The corresponding momentum during the magnetic cycle of the SPS in both cases is shown in Fig.~\ref{fig:mag-cycle}. In the case of the low energy beam injected on the collimator, the data were collected at the injection plateau during the first second of the cycle. Before the magnets started to ramp, the beam was dumped.

\begin{figure}
\centering
  \includegraphics[width=65mm,trim = 20mm 210mm 70mm 5mm, clip]{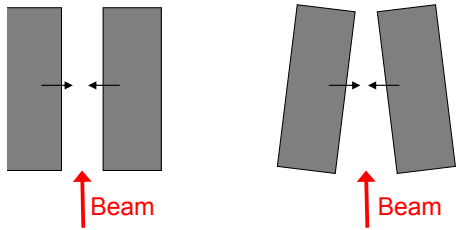}\\
  \caption{Schematic sketch of collimator jaws kept parallel to the beam direction when moved in (left) and with a 2~mrad angle (right).}\label{fig:jaw-angles}
\end{figure}

For the high energy measurements, the beam was instead accelerated up to 106.4~GeV/nucleon at the flat top, where the data-taking system continued to cycle while the magnets stayed at this energy and the RF system kept the beam at a constant energy. In this way the beam was circulating in the machine, while we scraped it with the collimator in several steps until all particles were lost. Because of the low intensity of the \pb beam, the collimator jaws had to be moved in close to the core of the beam (typical values are around 1~$\sigma$) in order to create significant loss signals. We performed measurements using two different angles of the collimator with respect to the beam axis (0~mrad and 2~mrad) as schematically illustrated in Fig.~\ref{fig:jaw-angles}, in order to vary the effective length travelled by particles in the collimator.

The circulating beam current was measured by a Beam Current Transformer (BCT). These measurements were used later to normalize the BLM signals by the number of lost particles. The BCT has a resolution of a few $10^8$ charges and the BLMs can detect around $10^7$ lost charges nearby. Since typical losses were a few $10^9$ charges, the combined uncertainty on the normalized signal is around 10\%. In the case where the beam was injected on the collimator, the normalization of the losses by the BCT could not be performed, since the injected intensity varied between different cycles.

For comparison, data were also taken during high-intensity proton runs with the beam circulating at the flat top in the same way as for \pb ions. These beam parameters are also presented in Tab.~\ref{tab:SPS-param}.

\section{SIMULATION TOOLS}
\label{sec:icosim}
In order to simulate the particle propagation through a lattice together with the particle-matter interactions in the collimators, a specialized program, ICOSIM (for Ion COllimation SIMulation), has been developed~\cite{braun04}. ICOSIM reads files from MAD-X~\cite{madx} defining the optics and aperture of a machine, in order to make it straightforward to simulate any machine for which such a representation exists. The program creates an initial beam distribution that is tracked in the variables $(x,x',y,y',\delta,A,Z)$ through the lattice. The number of particles in the distribution is not constant, since interactions in the collimators may split particles into several smaller fragments with different values of $A,Z$ and particles hitting the physical aperture are removed from the tracking. Synchrotron oscillations are neglected, because typical time scales are long ($T_{RF}\approx 500$~turns for the LHC and $T_{RF}\approx440$ turns for the SPS) compared to the time between a first interaction of an ion with a collimator and the final loss (1-10 turns for most particles, 100 turns in rare cases).

Particles are tracked up to the first collimator impact using a linear matrix formalism between collimators with a small artificial emittance blowup on every turn, which accounts for diffusion under single beam effects, which need not be specified. The linear tracking is typically done for 10$^5$~turns and serves to generate impact coordinates on the collimator. This method was developed with the aim of simulating fixed jaws and a growing envelope. We have chosen to use the same method to simulate the SPS experiments, although in this case the jaws were moving into the beam with a speed of 4~mm/s. This choice was made with the intention of benchmarking the specific generation of impact coordinates on the collimator and since the steps of the jaw were small compared to the discontinuities in the SPS aperture. As explained later, this fact makes the ratio between particles lost at different impact locations almost independent of the jaw positions. Thus, it is a very good approximation to use this simulation method also in the SPS. For the same reason, impact positions are relatively independent of orbit distortions and therefore we use the perfect machine optics in the tracking.
The diffusion of the envelope for the SPS simulations was chosen to 92~nm/turn, corresponding to the jaw speed. The resulting average impact parameter (defined as the distance between particle impact and the inner edge of the collimator jaw) is 22~$\mu$m.

From the first collimator impact, element by element tracking is used in ICOSIM and chromatic effects at leading order, as well as sextupoles in thin kick approximation, are included. Higher order multipoles are neglected, since particles are typically lost in less than 100~turns, which is not enough to see a significant effect of small non-linearities. At the end of each element, a check is performed to determine which particles are outside the aperture. The impact momenta and coordinates on the vacuum chamber are estimated by a linear interpolation inside the magnetic element.

Two different methods can be used to simulate the particle-matter interaction in the collimator:
\begin{itemize}
\item \textit{FLUKA XS}: ICOSIM has a simple built-in Monte-Carlo program, which includes multiple scattering in a Gaussian approximation~\cite{pdg}, ionization through the Bethe-Bloch formula, nuclear fragmentation and electromagnetic dissociation. The last two processes are simulated by sampling from tabulated cross sections, created by FLUKA for all possible fragments and fragmentation channels. Only the heaviest fragment created in each interaction is tracked. In the LHC, only these fragments are important to follow, since ions with $|\delta| > 0.05$ are already lost in the warm collimation insertion.

\item \textit{FLUKA full}: The transport and interaction of all particles through the collimator region is evaluated on each turn by external calls to FLUKA, which simulates the full shower in a 3D model of the collimator geometry. This method is more detailed and sophisticated but significantly slower in terms of computation time.
\end{itemize}

The BLM signals depend not only on the number of ions lost nearby, but also on the mass of the ions (assuming the same energy per nucleon, a heavier ion carries more energy), the distribution of impact parameters and the amount and type of material they have to traverse before reaching the monitor. At some BLMs, with less nearby material, particles lost far away may cause a signal, while BLMs that are well shielded by magnetic elements may only see small traces of the showers caused by the closest losses. In order to accurately simulate this for a quantitative comparison with data, the particle-matter interaction of the lost ions in the full geometry needs to be taken into account.

\begin{figure}[tb]
\centering
\includegraphics*[width=85mm,trim = 35mm 20mm 25mm 50mm, clip]{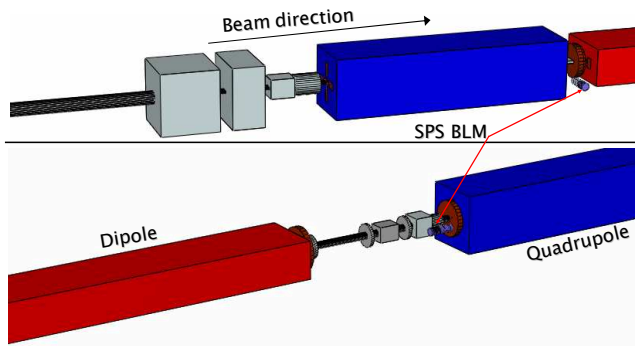}
\caption{The 3D geometry as implemented in FLUKA around a BLM in the vertical plane (BL520, top) and in the horizontal plane (BL521, bottom). A photo of the central part of the real geometry around BL521 is shown in Fig.~\ref{fig:sps-blm}.}
\label{fig:blm-geo-fluka}
\end{figure}

\begin{figure}[tb]
\centering
  \includegraphics[width=85mm]{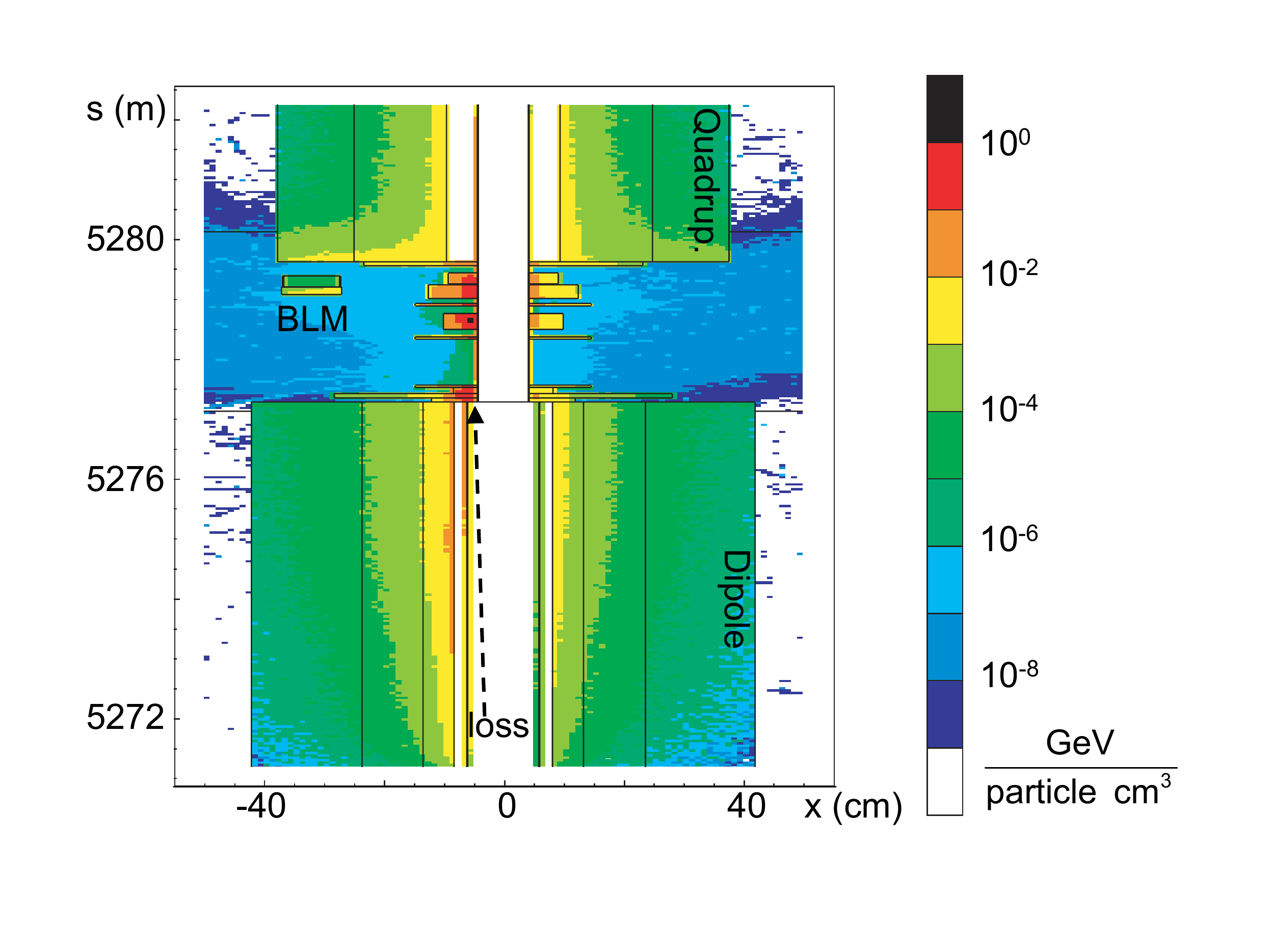}\\
  \caption{The geometry as implemented in FLUKA around the monitor BL521 with the simulated energy deposition from a typical loss superimposed. The loss is indicated by the dashed arrow.}\label{fig:shower}
\end{figure}

As discussed later, the main loss location is right downstream of the collimator. Thus, the 3D geometry of the magnetic elements around the monitors BL520, BL521, BL522 and BL523 was
implemented in FLUKA, as illustrated for BL520 and BL521 in Fig.~\ref{fig:blm-geo-fluka}. The magnetic field in the magnets nearby was neglected. A simulation including the magnetic field of the quadrupole nearest to BL520 showed a negligible difference with respect to the case without any field. The momenta and impact coordinates on the inside of the vacuum pipe of all particles lost within a 15~m interval of each BLM were recorded in ICOSIM and fed as starting conditions into FLUKA and the resulting energy deposition in the N$_2$ gas inside the BLMs was converted to dose in Gy. An example of the simulation of the shower caused by the lost particles close to BL521 is shown in Fig.~\ref{fig:shower}.

In the ICOSIM simulations of the SPS, typically more than $2\times10^5$ particles were tracked in order to arrive at a statistical uncertainty of 1--5\% on the final simulated BLM signal, depending on BLM.

\section{RESULTS: 106.4 {GeV}/NUCLEON, PARALLEL JAWS}

\begin{figure}[tb]
\centering
\includegraphics*[width=80mm,trim = 22mm 5mm 10mm 0mm, clip]{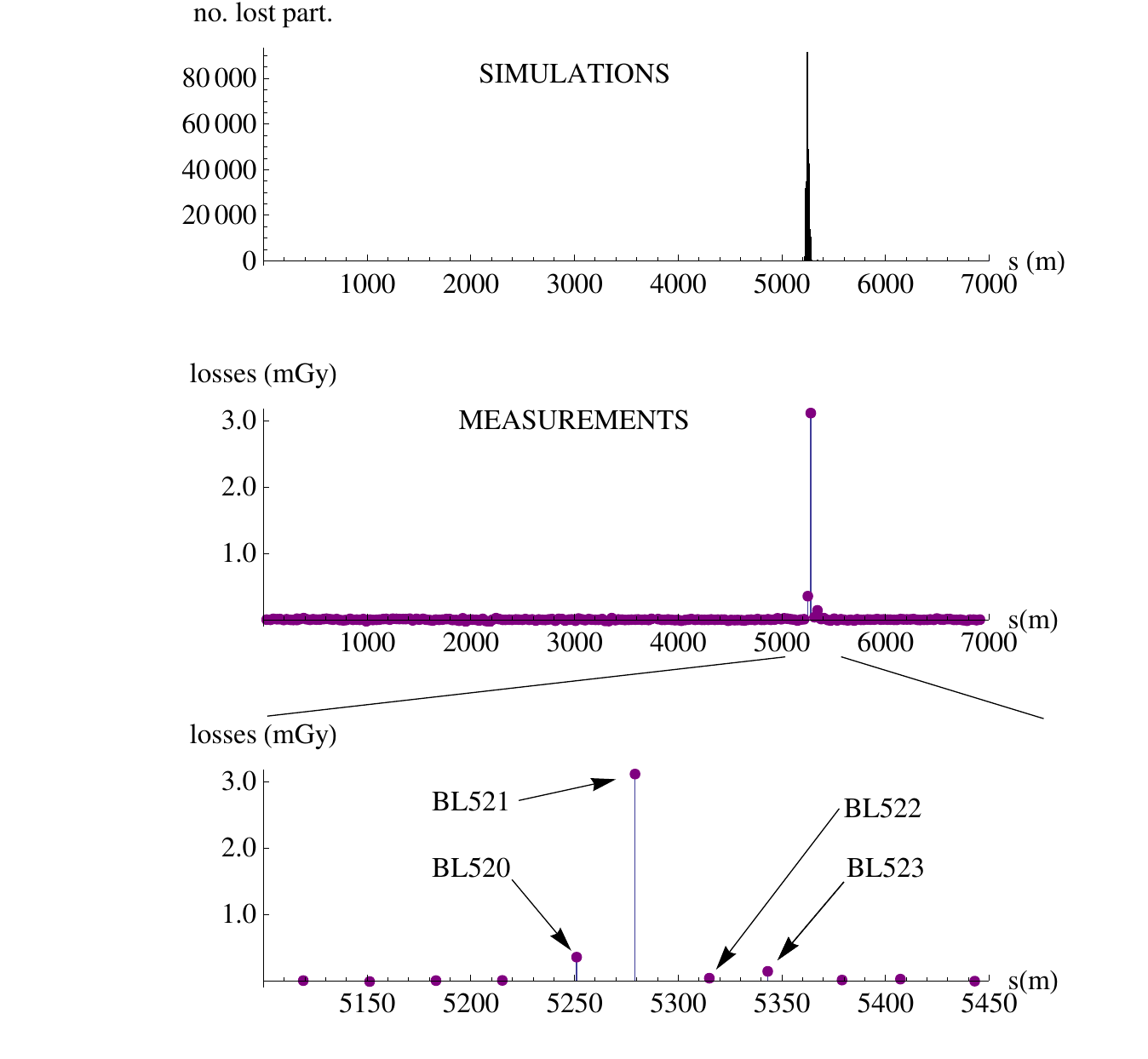}
   \caption{Example of simulated ICOSIM (top) and measured (middle) loss map of the whole SPS ring with a 106.4~GeV/nucleon circulating \pb beam. The bottom part shows a closeup of the loss peak in the measurements, with the names of the BLMs with the highest signals indicated.  The simulated losses were binned in 5~m intervals.  The collimator is located at $s=5222$~m, just upstream of the large loss peak.}
\label{fig:lm-ring-ions}
\end{figure}

\begin{figure}[tb]
\centering
\includegraphics*[width=80mm]{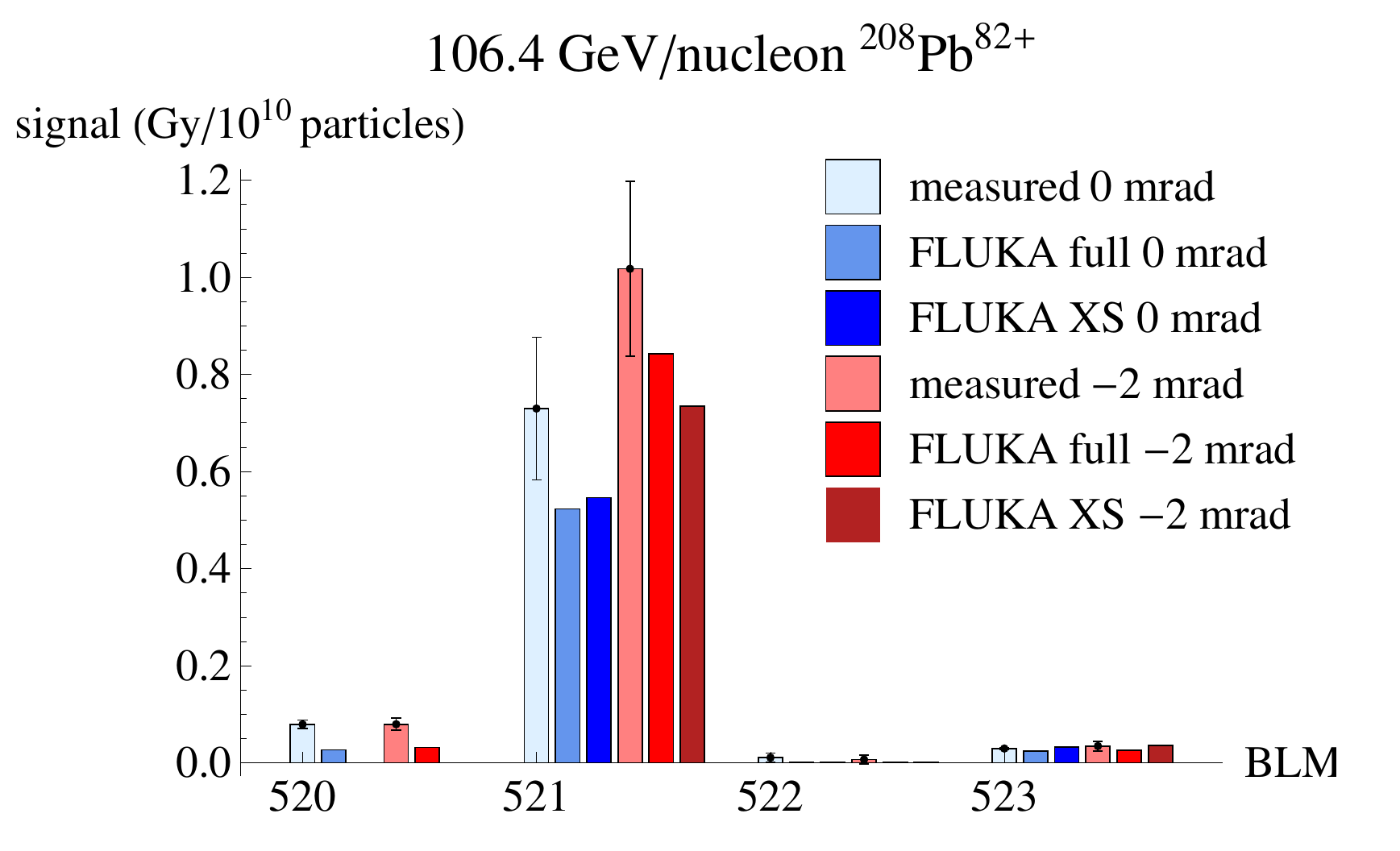}\\
\includegraphics*[width=80mm]{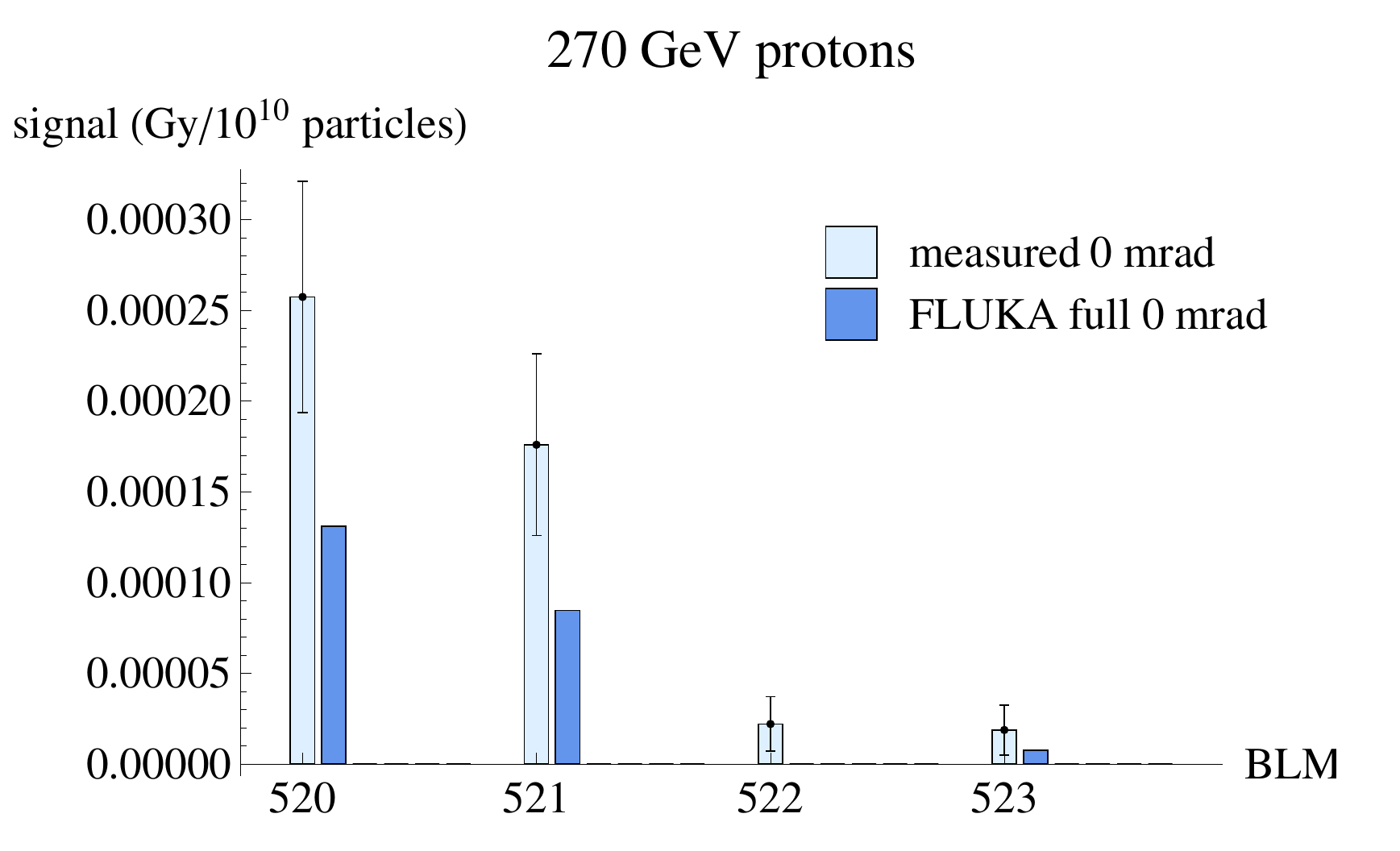}\\
\caption{Average loss map for 106.4~GeV/nucleon \pb ions (top) and 270~GeV protons (bottom) with parallel jaws (blue bars) and jaws tilted by 2~mrad (red bars). The measurements are compared with simulations using both methods of simulating the particle-matter interaction in the collimator, as described in text. All results were normalized to $10^{10}$ lost particles. The standard deviation between different measurements is indicated.}
\label{fig:lm-ions-blm}
\end{figure}

\begin{table} \centering
  \caption{The measured and simulated BLM signals (in Gy/$10^{10}$~lost particles) during 106.4~GeV \pb operation, as shown in the top part of Fig.~\ref{fig:lm-ions-blm}.}
  \label{tab:lm-ions-blm}
  \begin{ruledtabular}
\begin{tabular}{|l|r|r|r|r|}
  method/BLM & 520 & 521 & 522 & 523\\ \hline
  meas. 0~mrad & $0.079$ & $0.73$ & $0.01$& $0.030$\\
   & $\pm0.012$ & $\pm0.18$ & $\pm0.009$ & $\pm0.009$ \\ \hline
  \textit{FLUKA full} 0~mrad & 0.027 & 0.52 & 0.0008 & 0.024 \\ \hline
  \textit{FLUKA XS} 0~mrad & 0.0 & 0.55 & 0.0008 & 0.032 \\ \hline
  meas. -2~mrad & 0.079 & 1.02 & 0.007 & 0.034 \\
   & $\pm0.009$ & $\pm0.15$ & $\pm0.009$ & $\pm0.0019$ \\ \hline
  \textit{FLUKA full} 0~mrad & 0.031 & 0.84 & 0.0007 & 0.026 \\ \hline
  \textit{FLUKA XS} 0~mrad & 0.0 & 0.73 & 0.001 & 0.036 \\
\end{tabular}
\end{ruledtabular}
\end{table}

We focus first on the case of a 106.4~GeV/nucleon circulating \pb beam and parallel collimator jaws and compare with other cases in later sections.

A typical example of a loss map measured during a machine cycle with these conditions is shown in Fig.~\ref{fig:lm-ring-ions}, together with the corresponding simulated loss map from the \textit{FLUKA full} method. The detector background, consisting of noise and other beam losses not caused by the collimator movement, had to be subtracted. As background we used the loss map from the machine cycle before the collimator movement. A similar approach was already used in Ref.~\cite{redaelli06} to benchmark the SixTrack program~\cite{schmidt94} for proton beams. The only stable loss locations, clearly separable from the background, are just downstream of the collimator, both in simulations and measurements. This holds true also with jaws tilted by 2~mrad and at 5.9~GeV/nucleon. Therefore, in the remainder of this text we focus on the four BLMs downstream of the collimator which see the highest signal.

The top part of Fig.~\ref{fig:lm-ions-blm} and Tab.~\ref{tab:lm-ions-blm} show the average measured BLM signals, normalized to $10^{10}$ lost particles (using the BCT) and averaged over several machine cycles, together with the corresponding simulation results for the different methods of representing the particle-matter interactions in the collimator, which we discuss in Sec.~\ref{sec:sim-methods}. The ion loss pattern from the measurements is qualitatively very similar to the simulations, with the maximum signal on BL521.

\begin{figure}[tb]
  \includegraphics[width=75mm]{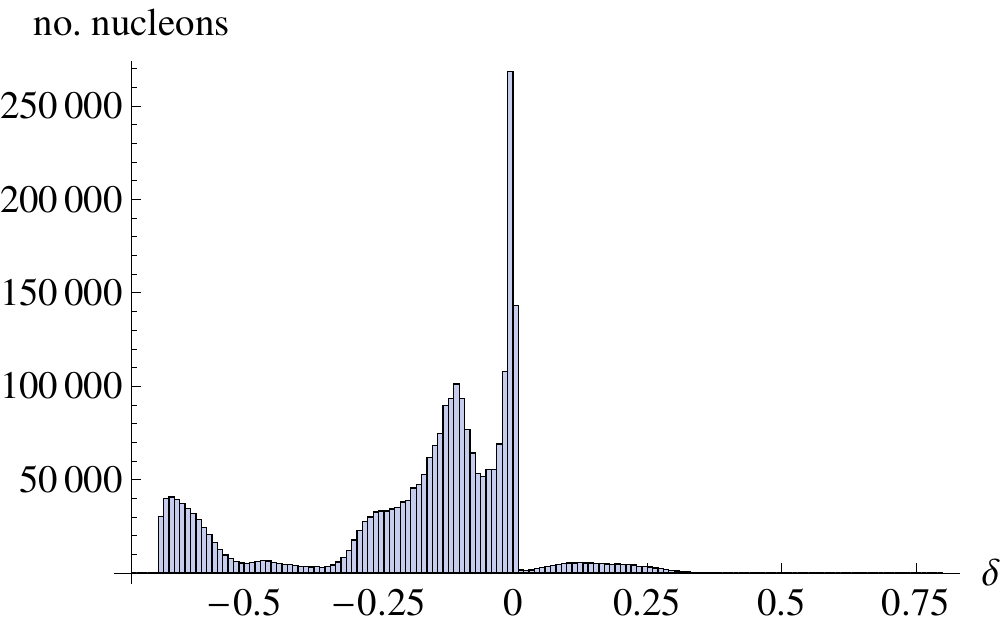}\\
  \includegraphics[width=75mm]{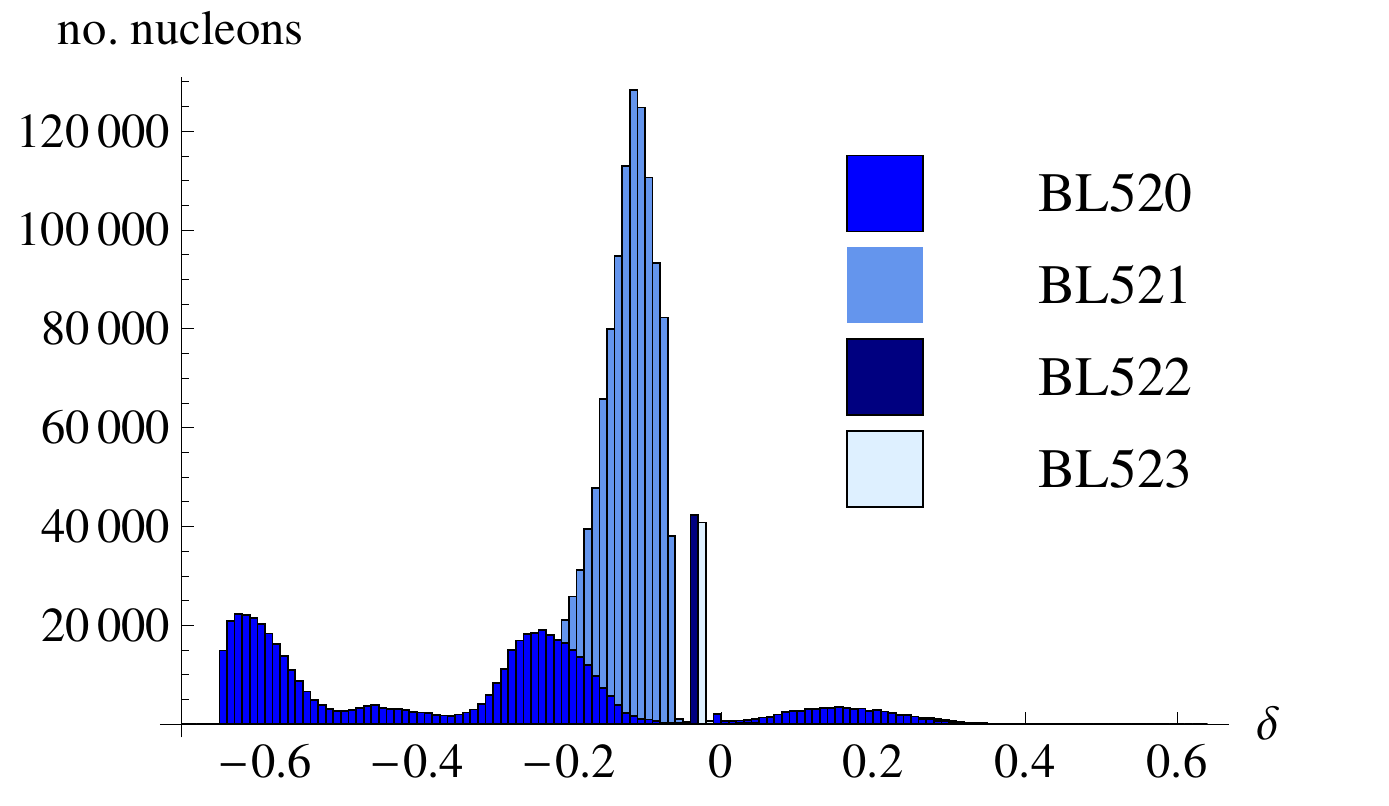}\\
  \caption{The simulated distribution of $\delta$ calculated according to Eq.~(\ref{eq:delta-eff}), using the \textit{FLUKA full} method, of all high energy nuclei except \pb coming out of the collimator (top) and of the ions lost within a 15~m interval upstream of each BLM (bottom) at 106.4~GeV/nucleon with parallel jaws. The height of the bars show the number of nucleons belonging to ions having $\delta$ within a certain interval.}
  \label{fig:delta-spectrum-coll}
\end{figure}

The fact that almost all losses occur close to the collimator, as well as the ratio between signals on different monitors, can be qualitatively understood by considering first the simulated $\delta$-spectrum of nuclei leaving the collimator, which is shown in Fig.~\ref{fig:delta-spectrum-coll}. We show the number of nucleons originating from ions with a certain $\delta$, since the FLUKA simulations show that an ion impinging in a magnetic element is fully fragmented within a few decimeters, meaning that the BLMs intercept mainly secondary particles originating from the hadronic shower, which to good approximation is similar to the shower created by the corresponding number of free nucleons~\cite{agosteo01}. Therefore the BLM signal caused by an ion lost in a specific location is approximately proportional to its number of nucleons. The heights of the peaks in the figures give thus an approximate estimate of the fraction of the BLM signal caused by different values of $\delta$.

\begin{figure}[tb]
\begin{center}
\includegraphics*[width=75mm]{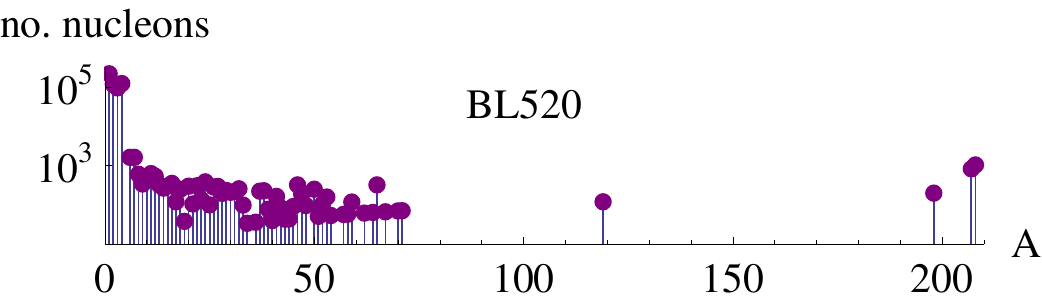}\\
\includegraphics*[width=75mm]{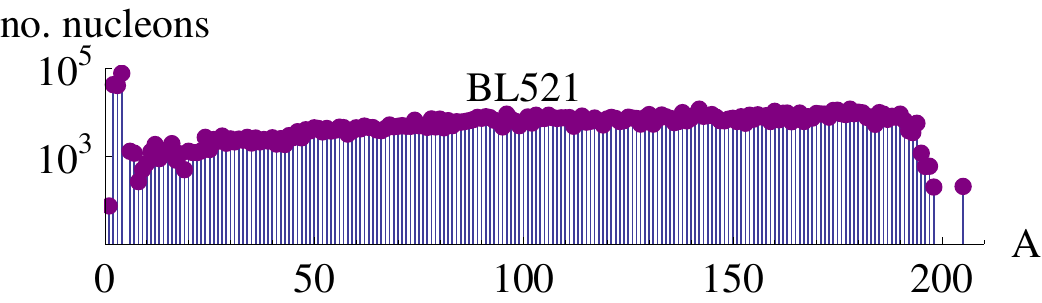}\\
\includegraphics*[width=75mm]{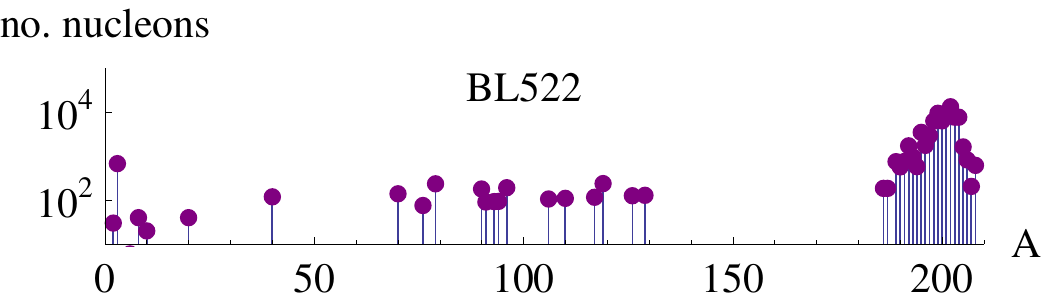}\\
\includegraphics*[width=75mm]{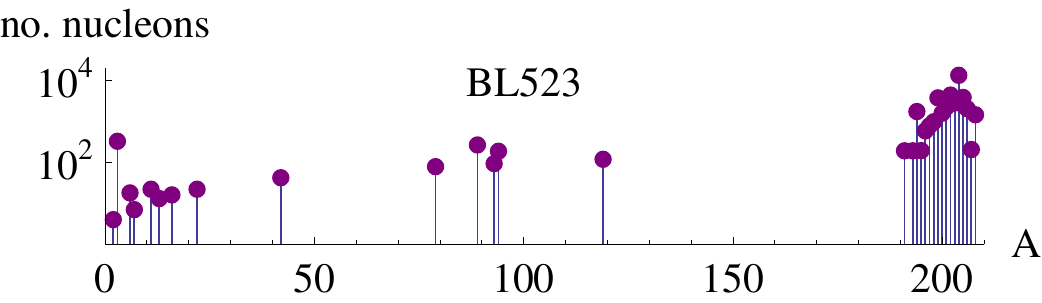}\\
\end{center}
  \vspace{-2 mm}
   \caption{The $A$ distribution of ion fragments lost within 15~m upstream
     of each BLM at 106.4~GeV/nucleon and parallel collimator jaws as simulated with the \textit{FLUKA full} method.  The heights of the bars show the number of nucleons from
     ions having a certain $A$.\vspace{-1mm}}
     \label{fig:A-spectrum-BLMs}
\end{figure}

\begin{figure}[tb]
  \includegraphics[width=75mm]{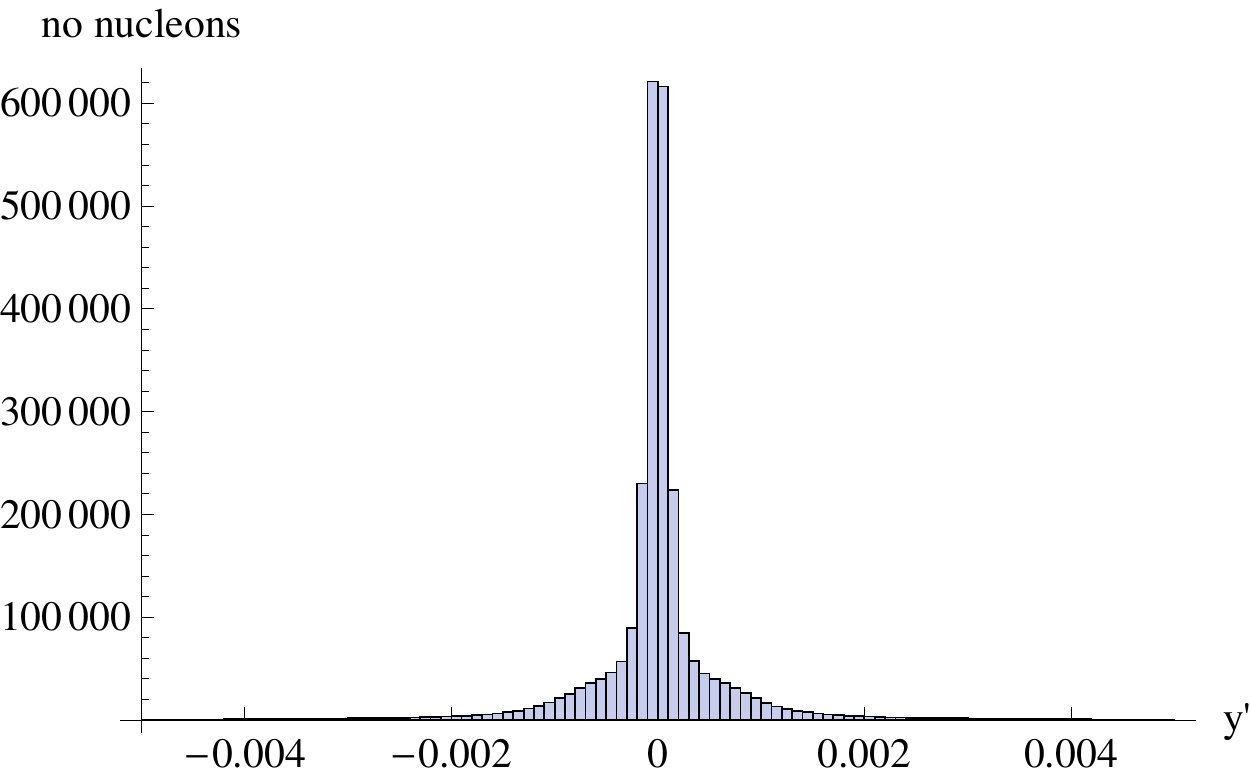}\\
  \includegraphics[width=75mm]{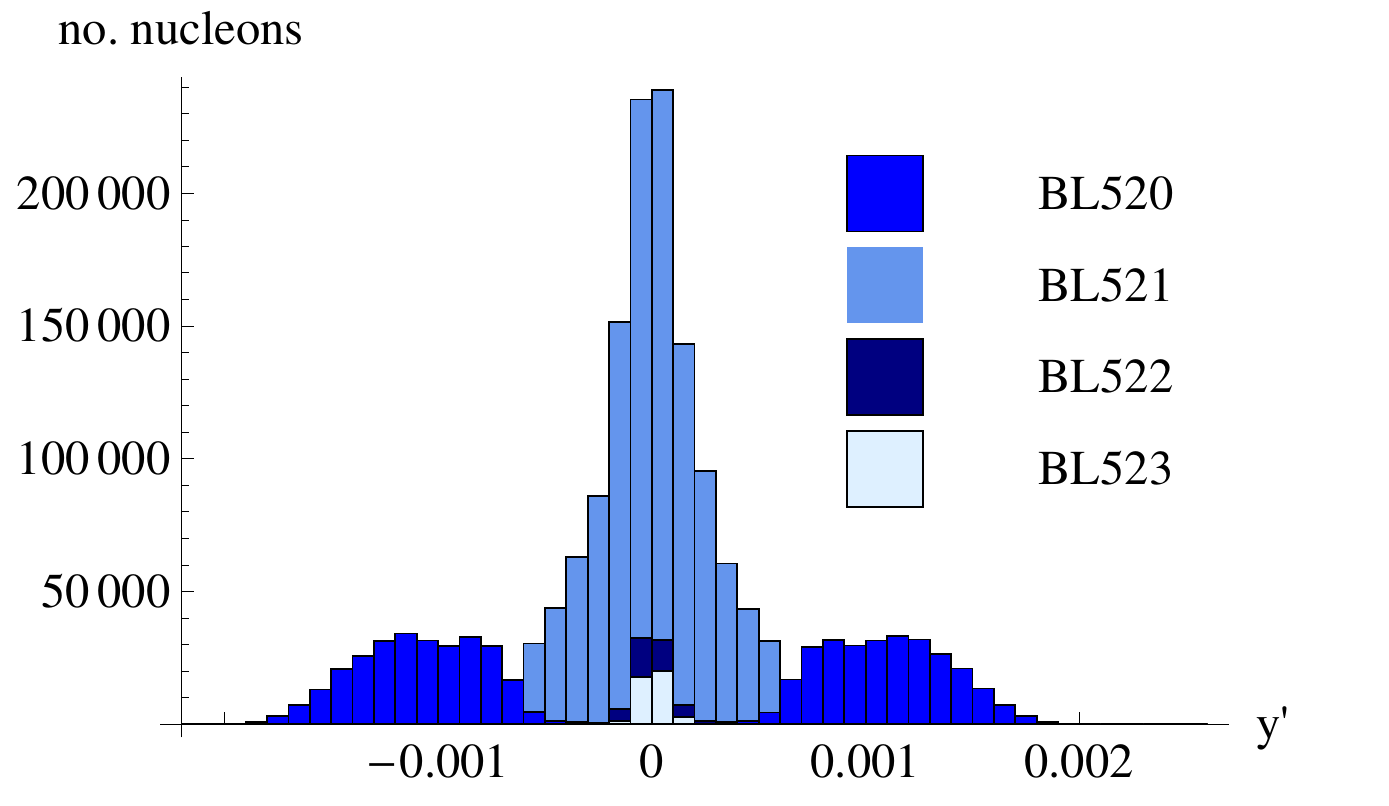}\\
  \caption{The simulated distribution from the \textit{FLUKA full} method of $y'$ of all ion fragments coming out of the collimator (top) and of the ions lost within a 15~m interval upstream of each BLM (bottom). The height of the bars show the number of nucleons belonging to ions within a certain interval.}
  \label{fig:py-spectrum-coll}
\end{figure}

In order to explain the loss pattern, we consider this $\delta$-spectrum in Fig.~\ref{fig:delta-spectrum-coll} together with the dispersive orbits of ion fragments with these $\delta$-values starting at one of the collimator jaws, shown in Fig.~\ref{fig:ions-disp}. Here several trajectories for typical values of $\delta$ are presented, together with the $s$-values of the BLMs. Fragments with $|\delta|<0.013$ stay inside the vacuum chamber and can make a full turn in the machine, while particles with larger $|\delta|$ are lost deterministically downstream of the collimator. Fig.~\ref{fig:ions-disp} clearly demonstrates the spectrometer effect discussed earlier, since fragments with different values of $\delta$ are lost at different longitudinal positions. Values of $\delta$ for some common fragments are shown in Tab.~\ref{tab:delta-fragments}. Therefore, each of the four BLMs considered in detail sees a different spectrum of ions, shown in Fig.~\ref{fig:A-spectrum-BLMs}. This can help us to understand the origin of the signals at each BLM:

\begin{itemize}
\item \textit{BL520}: The main loss mechanism in the vicinity is not dispersion but instead the vertical aperture, which intercepts particles with large betatron angles. This can be understood from the spectrum of vertical angles of the fragments exiting the collimator, as shown in Fig.~\ref{fig:py-spectrum-coll}, and typical vertical orbits of particles with large scattering angles as shown in Fig.~\ref{fig:vert-beta-orb}. The horizontal aperture is significantly larger (see Figs.~\ref{fig:ions-disp} and \ref{fig:vert-beta-orb}), making the vertical aperture the limitation for large-angle particles. Mainly light fragments are lost close to BL520, consistent with the expectation that they receive significantly larger transverse recoils in the fragmentation processes. The expected signal is significantly lower than measurements, maybe because BL520 is located only 30~m downstream of the collimator. It might see traces of shower particles from the collimator, in particular high energy neutrons created by electromagnetic dissociation, for which we have not attempted a detailed modelling. We have estimated this contribution through a FLUKA simulation of the neutron propagation through a rough model of the tunnel and around 45\% of the signal shown in Fig.~\ref{fig:lm-ions-blm} (top) on BL520 comes from showers induced by scattering neutrons. However, since the neutron contribution is very sensitive to objects in the tunnel acting as scatterers, which are not properly taken into account in the simulation, this should be considered as a rough estimate.

\item \textit{BL521:} Ions with $-0.2<\delta < -0.08$
are lost near the aperture limitation at $s=5277$~m, close to BL521. Fig.~\ref{fig:delta-spectrum-coll} demonstrates that this corresponds to a large fraction of the fragments and Fig.~\ref{fig:A-spectrum-BLMs} shows that BL521 sees a very wide spectrum of ions. Since BL521 is located only 2~m downstream of this position in negative $x$ with almost no shielding material in between (see Figs.~\ref{fig:sps-blm}, \ref{fig:blm-geo-fluka}, and~\ref{fig:shower}), this monitor is expected to show a very high signal when ion beams are scraped with the collimator. Consequently, BL521 has the maximum signal both in measurement and simulations, which predict a value around 30\% lower than measurements.

\item \textit{BL522}: BL522 is placed in the vertical plane, while the dispersive losses of mainly heavy fragments occur in the horizontal plane. This makes the uncertainty of the shower simulation much higher, since the signal is caused by secondary particles very far away from the central core of the shower. BL522 showed a much higher signal in measurements than simulations, although fluctuations between different cycles were very large (of the same order of magnitude as the measured signal), a fact that is not well
understood. BL522 had the lowest signal (around 1.4\% of the signal on BL521) of the BLMs that we consider in detail.

\item \textit{BL523}: Only particles with a magnetic rigidity ($\delta\approx -0.02$) close to the
original \pb\ ion are lost in the vicinity. Therefore the spectrum of lost ions consists mainly of heavy fragments. This situation is similar to what can be expected in the cold regions of the LHC, since all particles having $|\delta|\geq 0.05$ are already lost in the warm insertion. The signal from the \textit{FLUKA full} simulation is 17\% lower than measurements.
\end{itemize}

\begin{table} \centering
  \caption{The $\delta$ from fragmentation of the most common ions created in the collimator as calculated with Eq.~\ref{eq:delta-eff} for $\Delta p/p_0=0$.}
  \label{tab:delta-fragments}
  \begin{ruledtabular}
\begin{tabular}{|l|l|l|l|l|l|}
  $^{207}$Pb & $^{207}$Tl & $^{206}$Pb & $^{206}$Tl & $^{205}$Pb & $^{205}$Tl \\
  -0.0048 & 0.0075 & -0.0096 & 0.0026 & -0.014 & -0.0023 \\ \hline
  $^{204}$Hg & $^{4}$He & $^{3}$He & $^{3}$H & $^{2}$H & $^{1}$H \\
  0.0053 & -0.21 & -0.41 & 0.18 & -0.21 & -0.61 \\
\end{tabular}
\end{ruledtabular}
\end{table}

It is clear from Fig.~\ref{fig:ions-disp} that the expected distribution of losses is relatively independent of closed orbit distortions (which were not accounted for in the simulations) and the position of the collimator. Displacing the trajectories in the figure vertically by a few~mm does not significantly change their impact positions in $s$, given the large impact angles and the aperture of the SPS. Likewise, the dispersive orbits starting at the other jaw, located in negative $x$, have a similar longitudinal impact distribution. Thus, only the magnitudes of the measured signals, not their relative ratios, change when the collimator is moved in closer to the core of the beam and therefore intercepts more particles. This behavior was predicted by simulations and confirmed by measurements.

\begin{figure}
  \includegraphics[width=85mm]{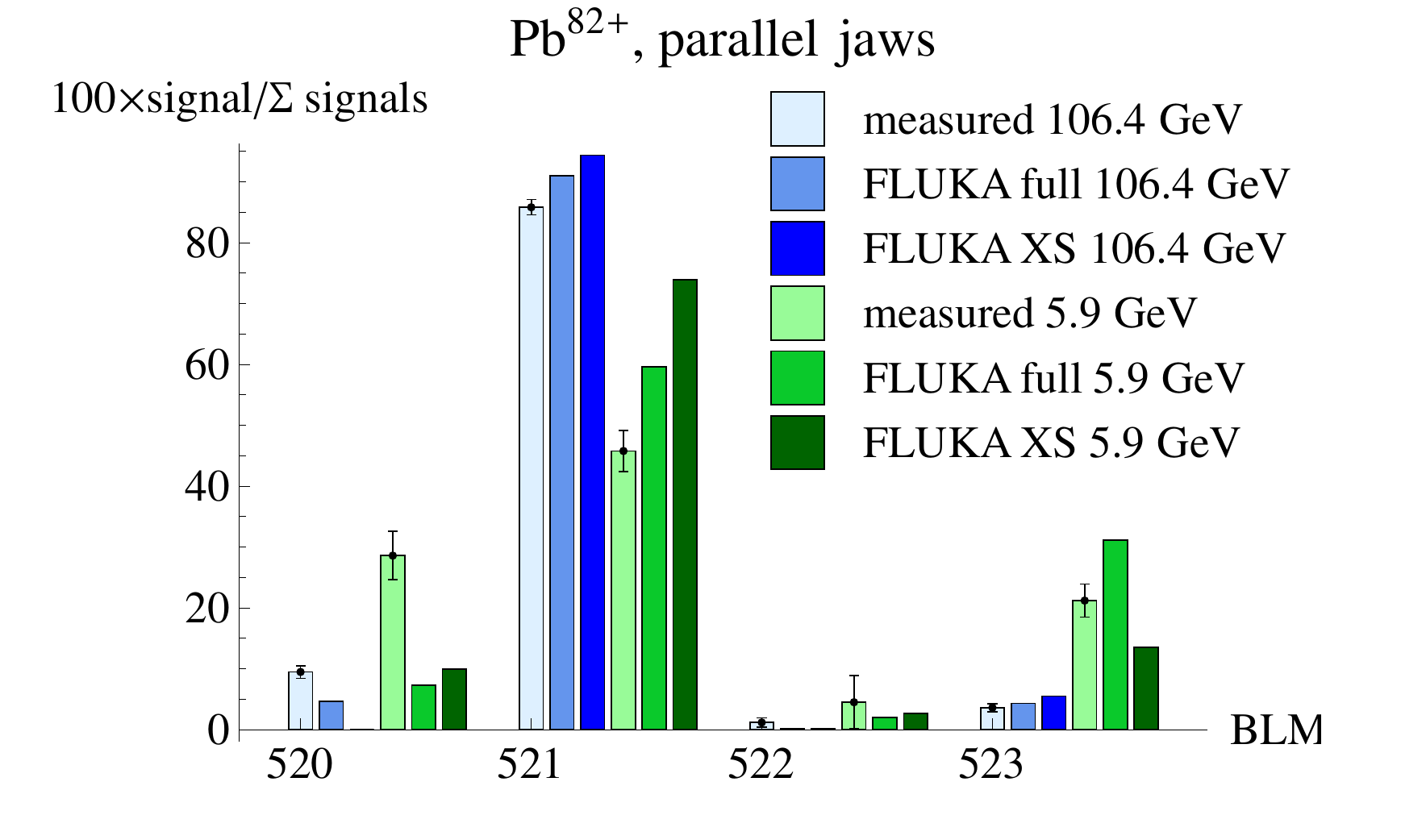}\\
  \caption{Measured and simulated \pb loss maps with parallel jaws for a 5.9~GeV/nucleon beam injected on the collimator (green bars) and circulating 106.4~GeV/nucleon (blue bars).}\label{fig:lm-ions-inj-ratio}
\end{figure}

The magnitudes of the simulated signals agree well with measurements (see Fig.~\ref{fig:lm-ions-blm}), although they are lower. To estimate the expected error, we assume approximately a factor 2-3 uncertainty of the FLUKA simulation of the energy deposition in the ionization chambers, since we consider secondary particles far from the core of a shower caused by ions with a grazing incidence. This is consistent with other comparisons between simulations and measurements of beam loss induced BLM signals at RHIC~\cite{prl07} and HERA~\cite{holzer07}.

Furthermore, the fragmentation cross sections in the collimator have significant uncertainties. A comparison between simulation codes done in a separate study, which we intend to publish elsewhere, suggests that the uncertainty of the largest cross sections to create specific isotopes may well be 50\%. Therefore, a new set of ICOSIM simulations were performed (using \textit{FLUKA XS}), where we in each run resampled all cross sections with a 50\% standard deviation around their initial value. This caused a standard deviation of 9\% of the signal on BL521 and approximately 20\% at the other BLMs, which see a narrower $\delta$-spectrum and therefore are more sensitive to variations in single cross sections.

There is also an uncertainty on the measured jaw angles, which has been estimated to be less than 0.2~mrad~\cite{braccoThesis08}. Further \textit{FLUKA XS} simulations show that this causes a 17\% deviation on the BLM signal. Other systematic errors in the measurement were estimated at 10\% (see Sec.~\ref{sec:exp-setup}).

However, these errors give only small corrections to the uncertainty of the FLUKA simulations. The same holds true for the single pass tracking from the collimator to the loss point, the influence of a non-perfect machine optics (as explained in Sec.~\ref{sec:icosim}) and for the impact coordinates on the collimator, which are much better known in the SPS than in the LHC. The SPS collimator was moved into the beam  close to the core, where the distribution is Gaussian with good approximation, while the LHC collimators will intercept halo particles, scattered to high amplitudes by beam dynamics processes that are hard to quantify.

Altogether, we expect the simulation to be accurate within a factor 3 and conclude that the discrepancies between measurements and simulations are within expected error margins, except for the highly fluctuating BL522. This error margin might be too pessimistic when considering the BLMs in the horizontal plane (BL521 and BL523).

\section{RESULTS: 106.4 {GeV}/NUCLEON, TILTED JAWS}
\label{sec:jaw-angle}
Fig.~\ref{fig:lm-ions-blm} shows the average measured and simulated loss maps with the collimator jaws tilted by 2~mrad when moved into the beam (see Fig.~\ref{fig:jaw-angles}). Compared to the case with parallel jaws the signal on BL521 increases by 40\% in measurement, which is well reproduced by the simulations (34\% in the \textit{FLUKA XS} simulation and 60\% in the \textit{FLUKA full} simulation). This increase can be understood from the fact that the average distance travelled by the particles inside the collimator decreases, meaning that less particles are absorbed.

Furthermore, the signal at BL523 shows a corresponding increase with tilted jaws both in measurements and simulations, while observed differences in the very small signal at BL522 are within the statistical error margin. At BL520 an increase by 18\% was predicted by the \textit{FLUKA full} simulation but not reproduced by the measurements.

\section{RESULTS AT 5.9 GeV/NUCLEON}

\begin{figure}[tb]
\begin{center}
\includegraphics*[width=75mm]{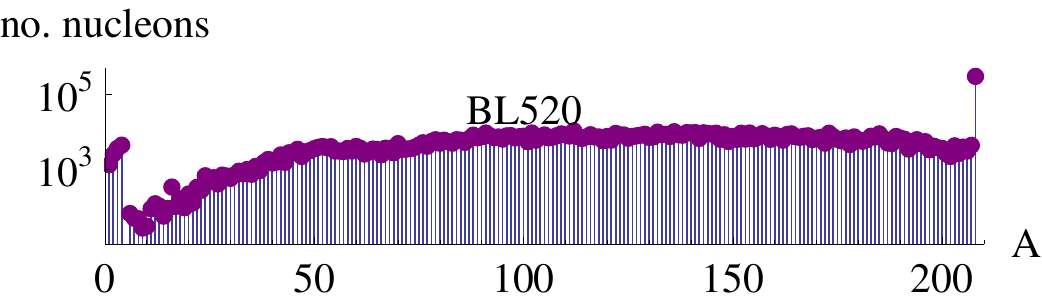}\\
\includegraphics*[width=75mm]{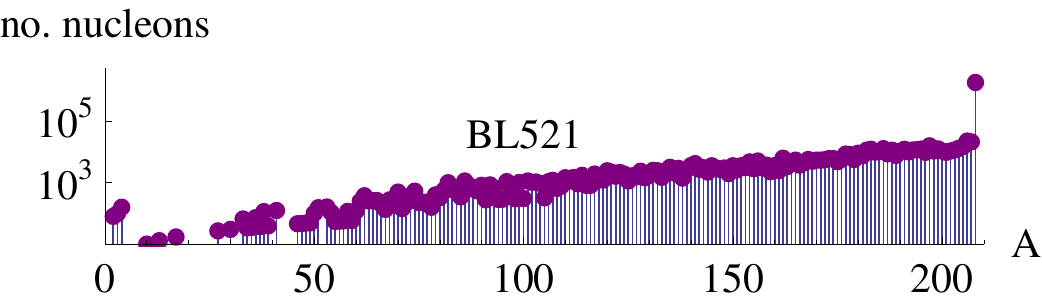}\\
\includegraphics*[width=75mm]{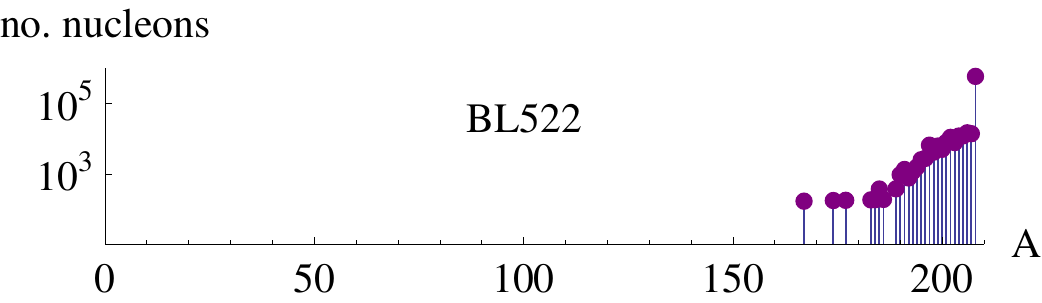}\\
\includegraphics*[width=75mm]{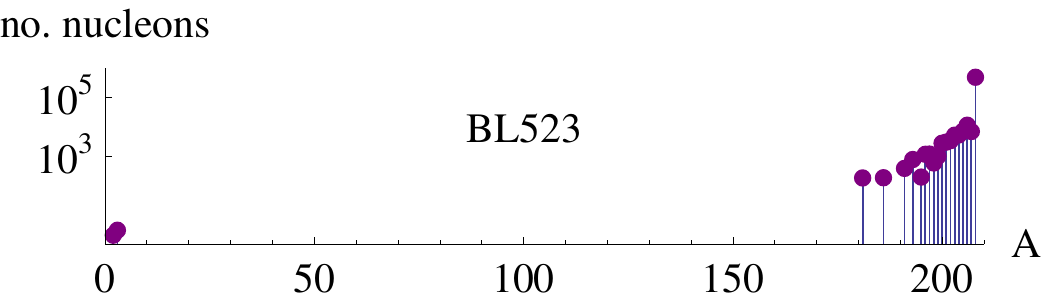}\\
\end{center}
  \vspace{-2 mm}
   \caption{The simulated $A$ distribution of ion fragments from the \textit{FLUKA full} method lost within 15~m upstream
     of each BLM at 5.9~GeV/nucleon.  The heights of the bars show the number of nucleons from
     ions having a certain $A$.\vspace{-1mm}}
     \label{fig:A-spectrum-BLMs-inj}
\end{figure}

\begin{figure}[tb]
  \includegraphics[width=75mm]{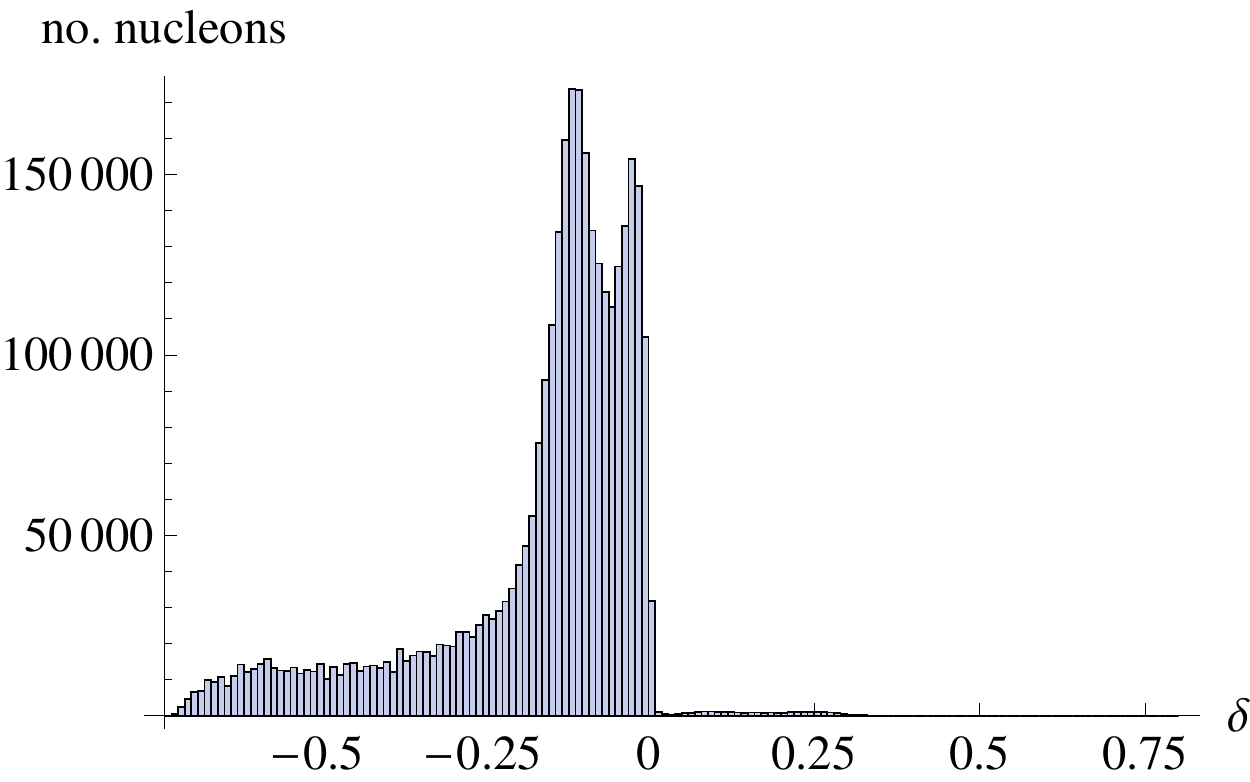}\\
  \caption{The simulated distribution of $\delta$ calculated according to Eq.~(\ref{eq:delta-eff}) from the \textit{FLUKA full} method of all ion fragments except \pb coming out of the collimator at 5.9~GeV/nucleon. The height of the bars show the number of nucleons belonging to ions having $\delta$ within a certain interval.}
  \label{fig:delta-spectrum-coll-inj}
\end{figure}

In Fig.~\ref{fig:lm-ions-inj-ratio} we show the measured and simulated loss maps using a 5.9~GeV/nucleon \pb beam injected on the collimator with parallel jaws, together with the results at 106.4~GeV/nucleon with parallel jaws presented earlier in Fig.~\ref{fig:lm-ions-blm}. Since the normalization to the intensity decay measured by the BCT was not possible (see Sec.~\ref{sec:exp-setup}), the normalization was instead done with respect to the sum of the four BLMs considered for both data sets and simulations. We noted however that the measured BLM signals dropped by around 5 orders of magnitude compared to the higher energy.

As shown in Fig.~\ref{fig:lm-ions-inj-ratio}, the fraction of the total signal at BL521 decreases at the lower energy, while it increases on BL520 and BL523. This is well reproduced by both simulation methods, although they consequently predict too low a fraction on the BLMs in the vertical plane (BL520 and BL522). The \textit{FLUKA full} method also shows lower magnitudes of the signals than \textit{FLUKA XS}.

The change in loss pattern can be qualitatively understood from the energy loss in the collimator. Using the Bethe-Bloch formula, we see that a 106.4~GeV/nucleon \pb ion loses around 0.16~GeV/cm in carbon due to ionization, while a 5.9~$A$~GeV \pb ion loses 0.13~GeV/cm. This corresponds however to very different fractional energy losses: 0.15\%/cm of the total energy for 106.4~$A$~GeV ions and 1.9\%/cm at 5.9~$A$~GeV. Since typical paths inside a collimator are several centimeters, the energy loss caused by ionization becomes significant at low energy (see the dispersive orbits in Fig.~\ref{fig:ions-disp}), meaning that the dispersion of the ions exiting the collimator is not only due to a different magnetic rigidity, but also because of a different energy per nucleon. The consequence is that most ions receive larger values of $|\delta|$ than at the higher energy and are lost further upstream in the ring. Furthermore, \pb ions that pass the collimator without fragmenting may receive a large enough $|\delta|$ to make them lost close to all considered BLMs. This can be seen in Fig.~\ref{fig:A-spectrum-BLMs-inj}, showing the mass spectrum of ions lost near each BLM.

At the same time the production of fragments changes slightly: At 5.9~GeV/nucleon, the cross section for electromagnetic dissociation goes down significantly~\cite{bert02}, while the cross section for nuclear inelastic interaction, on the other hand, is only weakly dependent on the energy~\cite{summerer90,summerer00}. Because of the different production rates and the ionization energy loss, the $\delta$-spectrum of the ions leaving the collimator is different at 5.9~GeV/nucleon. This is shown in Fig.~\ref{fig:delta-spectrum-coll-inj}.

\section{COMPARISON OF SIMULATION METHODS}
\label{sec:sim-methods}
In the \textit{FLUKA XS} simulations no signal is seen at BL520 with the 106.4~GeV/nucleon beam, since only the heaviest fragment produced in each collimator interaction is tracked, while the signal at BL520 is caused by very light fragments (see Fig.~\ref{fig:A-spectrum-BLMs}). During operation at 5.9~GeV/nucleon, heavy fragments are lost also at BL520 (see Fig.~\ref{fig:A-spectrum-BLMs-inj}), and consequently a non-zero signal is simulated by the \textit{FLUKA XS} method.

Furthermore, the \textit{FLUKA full} simulation at 106.4~GeV/nucleon predicts some smaller loss peaks between BL520 and BL521, caused by very light particles with a positive energy offset. These peaks are not present in the \textit{FLUKA XS} simulation but are also too far from the BLMs to be seen in the measurements.

However, at BL521, BL522 and BL523, where medium and heavy fragments are important, the \textit{FLUKA XS} method gives an excellent agreement both for parallel and tilted jaws, while the gain in tracking speed is on the order of a factor 10 compared to the \textit{FLUKA full} method. Furthermore, \textit{FLUKA XS} reproduces well the qualitative changes of the loss pattern when the beam energy is changed. Therefore, we conclude that \textit{FLUKA XS} is the preferred method for large machines with many collimators where light fragments are of low importance. This is the case for the LHC, since almost all light fragments are already lost in the warm regions.

\section{COMPARISON WITH PROTONS}
\label{sec:protons}

\begin{figure}[tb]
  \includegraphics[width=75mm]{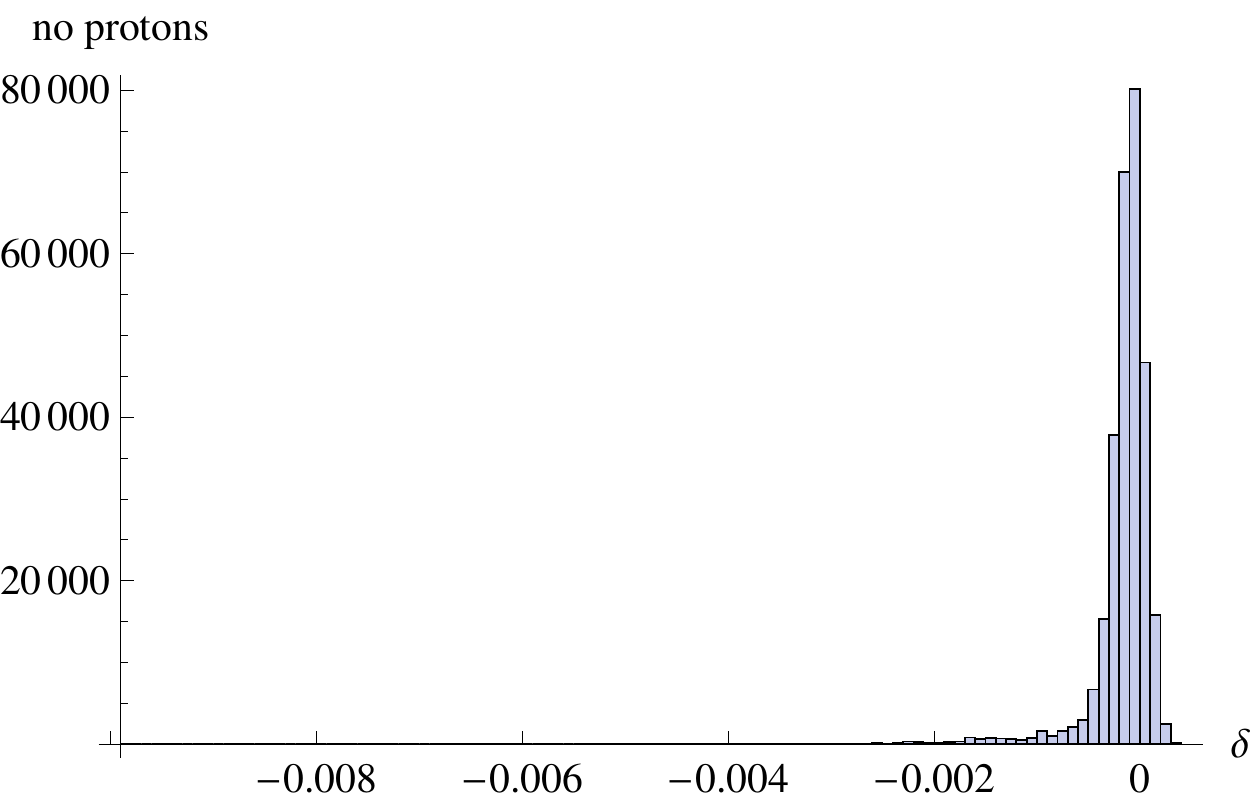}\\
  \includegraphics[width=75mm]{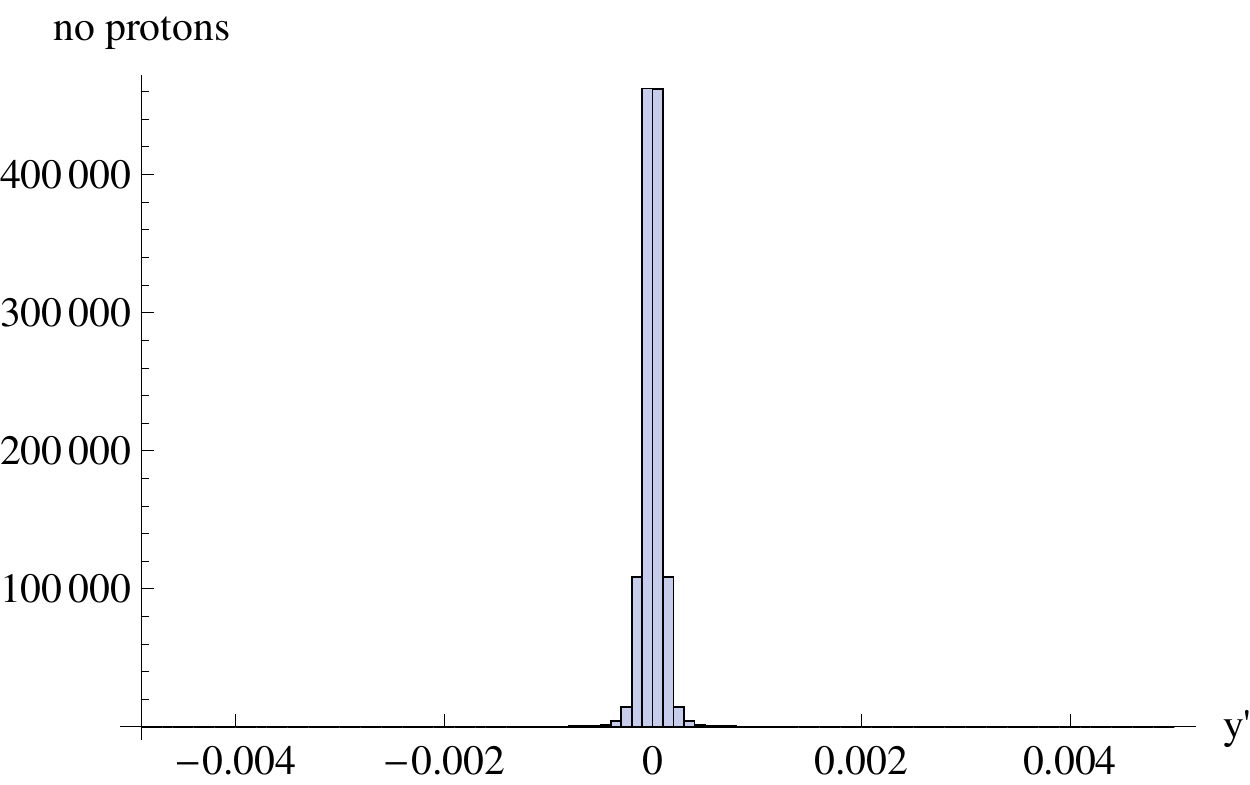}\\
  \caption{The simulated distribution of $\delta$ (top) and $y'$ (bottom) of all protons coming out of the collimator.}
  \label{fig:delta-spectrum-protons}
\end{figure}

For the sake of comparison, proton runs in the SPS were also simulated with ICOSIM, using the \textit{FLUKA full} method, and compared with measurements from September 2007 with a coasting high-intensity beam (parameters are given in Tab.~\ref{tab:SPS-param}). More details and analyzes of these measurements will be published elsewhere.

Average measured and simulated loss map are shown in Fig.~\ref{fig:lm-ions-blm}. The ratio between the higher peaks agrees well with measurements, although there is a factor 2 discrepancy in magnitude. This could, just as for ions, be explained with systematic errors dominated by the uncertainty in the shower simulation. Furthermore, the intensity was more than 4~orders of magnitude higher than in the ion runs, which might introduce additional losses not caused by the collimator. This is however not understood in detail.

It is clear from Fig.~\ref{fig:lm-ions-blm} that there is a significant qualitative difference between proton and ion loss patterns: the maximum signal for protons was found on BL520 (closest to the collimator), while in the ion runs it was found on BL521. This can be understood from the fact that the $\delta$ of the protons is much lower, since they cannot fragment, which can be seen from Fig.~\ref{fig:delta-spectrum-protons}. This means that large betatron angles caused by multiple scattering are the main loss mechanism instead of dispersion, although the spectrum of vertical angles is also more narrow than for ions (see Fig.~\ref{fig:delta-spectrum-protons}). The difference in loss pattern is a striking parallel to the expected behavior in the LHC, and the ability of the simulations to quantitatively predict this behavior in the SPS provides a very valuable benchmark of the simulation studies which have been performed for the LHC.

\section{CONCLUSIONS}
We have done measurements and simulations of \pb ion loss patterns induced by a collimator in the CERN SPS, in order to benchmark simulation tools used for the LHC. Qualitatively, the features of the loss pattern can be well understood from the particle-matter interactions in the collimator and the behavior of the dispersion function downstream of it. A wide range of nuclei are created in the collimator due to fragmentation of the original \pb beam. The created isotopes have different $Z/A$ ratios and follow different dispersive orbits until they are lost, making the machine act as a spectrometer.

Quantitatively, the simulated loss distribution corresponds well to measurements. In terms of absolute BLM signals, predicted and measured values agree to about 30\% for the largest losses, while discrepancies can reach a factor 3 for lower loss peaks (excluding the highly fluctuating BL522). This is consistent with expected uncertainties. We performed measurements with different beam energies and collimator angles and observed changes in the loss pattern, which were reproduced by our two-stage simulations with ICOSIM and FLUKA. We used two methods to simulate the particle-matter interaction in the collimator---a fast simplified Monte Carlo and a  full FLUKA shower simulation---and found a good agreement with measurements for both methods when the losses are dominated by heavy fragments close to the original ion. The simplified model is therefore well suited to predict limiting losses in the super-conducting magnets of the LHC with good accuracy. Losses consisting of light ions are on the other hand not treated by the simplified method, and if such losses are important the full shower simulation should be used.

Furthermore, we have studied beam loss data taken with a proton beam and found that the loss patterns induced by the collimator in the SPS are qualitatively different for \pb ions and protons. This difference is well reproduced by simulations. The ions are lost mainly due to a change in magnetic rigidity caused by fragmentation, while the protons are lost because of large betatron angles from scattering.

These results confirm and strengthen our understanding of ion beam losses related to collimation and are a vital test of our ability to make predictions for the LHC.

\section{ACKNOWLEDGEMENTS}
We would like to thank A.~Ferrari, M.~Magistris, G.I.~Smirnov and V.~Vlachoudis for inspiring discussions and help with FLUKA. Furthermore, we are thankful to M.~Stockner for discussions and for providing the technical specifications and characteristics of the BLMs, and to G. Arduini, B.~Dehning, M.~Jonker, D.~Kramer, D. Manglunki, C.~Zamantzas and the AB/OP group for their help during the measurements. We thank also L.~Jensen for providing specifications of the instrumentation.


\end{document}